# From Haloes to Galaxies - I

# The dynamics of the gas regulator model and the implied cosmic sSFR-history

Ying-jie Peng[1,2], Roberto Maiolino[1,2]

1 Cavendish Laboratory, University of Cambridge, 19 J. J. Thomson Avenue, Cambridge CB3 0HE, UK
2 Kavli Institute for Cosmology, University of Cambridge, Madingley Road, Cambridge CB3 0HA, UK

**ABSTRACT**

We explore the basic parameters that drive the evolution of the fundamental properties of star forming galaxies within the "gas regulator model", or bathtub-model. From the five basic equations of the typical gas regulator model, we derive the general analytic form of the evolution of the key galaxy properties, i.e. gas mass, star formation rate (SFR), stellar mass, specific SFR (sSFR), gas fraction, gas phase metallicity and stellar metallicity, without assuming that galaxies live in the equilibrium state. We find that the timescale required to reach equilibrium, $\tau_{eq}$, which is determined by the product of star-formation efficiency $\varepsilon$ and mass-loading factor $\lambda$, is the central parameter in the gas regulator model that is essentially in control of the evolution of all key galaxy properties. The scatters in most of the key scaling relations, such as the stellar mass-SFR relation and stellar mass-metallicity relation, are primarily governed by $\tau_{eq}$. Most strikingly, the predicted sSFR evolution is controlled solely by $\tau_{eq}$ (apart from the cosmic time), independent of the gas inflow rate and of the individual values of $\varepsilon$ and $\lambda$. Although the precise evolution of the sSFR depends on $\tau_{eq}$, the sSFR history is largely insensitive to different values of $\tau_{eq}$. The difference between the minimum and maximum sSFR at any epoch is less than a factor of four for any given values of $\tau_{eq}$. The shape of the predicted sSFR history simply mimics that of the specific mass increase rate of the dark matter halos ($sMIR_{DM}$) with the typical value of the sSFR around $2*sMIR_{DM}$. We show that the predicted sSFR from the gas regulator model is in good agreement with the predictions from typical Semi-Analytic Models (SAMs), but both are fundamentally different from the observed sSFR history. This clearly implies that some key process is missing in both typical SAMs and gas regulator model, and we hint at some possible culprit. We emphasize the critical role of $\tau_{eq}$ in controlling the evolution of the galaxy population, especially for gas rich low mass galaxies and dwarf galaxies that are very unlikely to live around the equilibrium state at any epoch and this has been largely ignored in many similar studies.

## 1. INTRODUCTION

Understanding galaxy formation and evolution is one of the most important issues in modern cosmology. The cosmological framework is well established and dark matter simulations of large-scale structure have been performed with great success. In order to produce realistic galaxies, baryon physics must be added onto the framework of dark matter haloes. However, due to the complexity of baryon physics such as star formation and feedback, these simulations usually fail to reproduce many of the observed properties of galaxies and also cannot clearly establish the relative importance of different processes in controlling the evolution of galaxy populations.

Observationally, new technologies and more powerful telescopes have enabled the observation of galaxies out to $z > 7$. Recent large multi-wavelength galaxies surveys such as the Sloan Digital Sky Survey (SDSS; York et al. 2000) locally, GAMA (Driver et al. 2009; Baldry et al. 2010), VIPERS (Guzzo et al., 2013), GOODS (Giavalisco et al. 2004), DEEP (Vogt et al. 2005; Weiner et al. 2005), DEEP2 (Davis et al. 2003), VVDS (Le Fèvre et al. 2005), COSMOS and zCOSMOS (Scoville et al. 2007; Lilly et al. 2007) and other deep surveys at high redshift, are delivering an unprecedented wealth of high quality data, which enable detailed studies of various galaxy properties and their evolution over a broad range of the cosmic time. This makes the fully empirical and phenomenological approaches become possible.

Although the heterogeneous population of galaxies appears complex, as it appears to be composed of largely different types and properties at first sight, when large samples of galaxies are studied, it appears that the majority of galaxies just follow simple scaling relations while the outliers represent some minority. In Peng et al. (2010) and (2012) we demonstrate the astonishing underlying simplicity of the galaxy population emerged from large surveys and derive the analytical forms for the dominant evolutionary processes that control galaxy evolution through continuity equations. The strategy is to use the observational material as directly as possible in order to identify the simplest things that are apparently demanded by the data and to define empirically based "laws" for the evolution of the galaxy population.

This simple model (hereafter P10-model) has successfully explained the origin of the Schechter form of the stellar mass function and reproduced many observed essential features of the evolving galaxy population over cosmic time. It has established a simple and self-consistent analytical framework to describe the stellar component of the evolving galaxy population, however, its connections to the cosmological framework of ΛCDM paradigm and to the gas content of the galaxy population are completely missed out in this model. This is because by design the P10-model is based entirely on observations. Although the properties of the dark matter haloes can be probed via gravitational lensing from imaging surveys or via clustering, and the gas content can also be observed through far-infrared (FIR) observations from Herschel and sub-millimeter imaging with ALMA, for the time being the accuracy and limited statistics of these surveys is not yet adequate to allow a similar approach of "reverse-engineering" as on P10.

The assembly history of dark matter haloes is well constrained both theoretically and via N-body simulations; the stellar component of the galaxy population can be nicely described by the P10-model; while the gas content is less constrained both theoretically and observationally, yet it is extremely important as gas is the fuel for star formation. Therefore, the goal of our next step is to explore and establish the crucial connections between the stellar component, gas component and dark matter halo component of the evolving galaxy population. In other words, implement the P10-model into the cosmological context.

It should be noted that the underlying philosophy of this new approach is very different from the usual theoretical approach of the semi-analytic models (SAMs), in the sense that the logical flow is reversed. SAMs start with the hierarchical build-up of dark matter haloes, within which baryons evolve subject to a large number of assumed physical processes to produce the stellar population of the galaxies. In order to reproduce observations, SAMs usually end up involving a large number of parameters and additional assumptions that are buried in the implementation. The inevitable complexity of the SAMs means that there is considerable uncertainty and degeneracy as to the uniqueness of any particular implementation, and the agreement with observations is never perfect.

Our new approach steps back from physical inputs, apart from the most basic continuity equations, and starts from the P10-model which is built entirely from observations and is required by the data. Then we try to implement the P10-model into the cosmological context via the gas regulator model as presented in Lilly et al. (2013) by identifying the main physical mechanisms that are actually responsible for controlling the evolution of the galaxy population. As stressed before, this type of new phenomenological approach becomes only possible because of the high quality data obtained from recent large surveys, which enables detailed or even precise study of various galaxy properties for the first time. The difficulty of this observation-based phenomenological approach is that it is difficult to establish the causal link between different observed quantities and the link between observations and theories.

In Section 2, we first introduce the typical gas regulator model. Then we derive the general analytic forms of the evolution of galaxy key properties and explore the dynamics of the gas regulator model. In Section 3, we test these analytic solutions by comparing them with the exact numeric solutions and show their dynamical evolution in different scenarios. In Section 4 we highlight the action of the equilibrium timescale $\tau_{eq}$ in regulating the sSFR evolution and discuss the difficulty of the gas regulator model to reproduce the observed sSFR history. In Section 5 we discuss and emphasize the critical role of the equilibrium timescale $\tau_{eq}$ in controlling the evolution of galaxy population. In Section 6, we summarize our findings. The cosmological model used in this paper is a concordance ΛCDM cosmology with $H_0 = 70$ kms$^{-1}$Mpc$^{-1}$, $\Omega_\Lambda = 0.75$ and $\Omega_M = 0.25$. Throughout the paper, we use the term "dex" to mean the antilogarithm, i.e., 0.1 dex = $10^{0.1}$ = 1.258.

## 2. THE GAS REGULATOR MODEL

The "gas-regulator" model (Lilly et al. 2013) or more commonly called as "bathtub" model (Bouché et al. 2010), and to a wider extent the bathtub-type models (e.g. Finlator et al. 2008, Recchi et al. 2008; Davé et al. 2012; Dayal et al. 2013; Dekel et al. 2013; Feldmann et al. 2013; Pipino et al. 2014; Dekel et al. 2014), generally takes into account the basic key physical processes of inflow, outflow, star formation and metal production. The formation of stars is instantaneously regulated by the mass of gas reservoir through the efficiency of start formation (i.e. the Schmidt-Kennicutt law) and through the mass-loss scaling with the SFR. The gas regulator model offers a simple way to link together the mass assembly of the dark matter haloes, the evolution of the gas content, metal content and stellar population of the galaxies through cosmic time. The gas regulator model has qualitatively successfully reproduced many key features of the galaxy population, such as the mass-metallicity relation (Lequeux et al. 1979; Tremonti et al. 2004), the fundamental metallicity relation (FMR, Mannucci et al. 2010) and can be employed to interpret observations like the metallicity dependence on environment (Peng et al. 2014). In some sense the gas regulator model, as a simple toy type model, can be regarded as a simplified version of the full Semi-Analytic Model (SAM) treatment (e.g. White & Frenk 1991; Baugh 2006 for a review), aiming to understand some specific physical processes and their dynamical behaviors in galaxy formation and evolution in a fully analytic way.

Recently the gas regulator model approach has been criticized for incapable of quantifying higher-order effects such as the scatter in individual scaling relations and various fundamental metallicity relations, for using the assumption of equilibrium state (e.g. Forbes et al. 2013). It is true that many of the current gas regulator models are based on the presumption that galaxies live roughly around the equilibrium state and little attention has been paid to the galaxies that are out of equilibrium. Indeed, we will show in the current paper (and in our future work) that the actual timescales for galaxies to achieve the equilibrium state are critical in determining many key properties of the galaxy population. We stress that the assumption of equilibrium is not an inherent feature or a requirement in the gas regulator model approach.

We will show that the gas regulator model approach is very powerful to study galaxies that live roughly in equilibrium state and those that are completely out of equilibrium as well. In the former case, the galaxies properties are mainly determined by the equilibrium values. In the latter case, the galaxies properties are also critically dependent on the timescale to reach the equilibrium state.

In Section 2.1 we introduce the typical gas regulator model and its configurations, which is largely based on the Lilly et al. (2013) implementation and we follow the same notations as those in Lilly et al. (2013). In Section 2.2 we derive the general analytic solutions of the key galaxy properties in the gas regulator model. In Section 2.3 we discuss the dynamical behavior of the key galaxy properties and their evolution with time.

**2.1 Model Implementation**

(a) *DM halo accretion*

We define the average specific accretion rate, or specific mass increase rate, of DM haloes, $sMIR_{DM}$ as

$$sMIR_{DM} = \frac{1}{M_{halo}} \frac{dM_{halo}}{dt} \quad (1)$$

The accretion here includes all dark matter mass, gas mass and stellar mass. Also we do not differentiate between smooth accretion and clumpy accretion, i.e. mergers (with different mass ratio). We adopt the average $sMIR_{DM}$ derived from the cosmological hydrodynamic simulations by Faucher-Giguere et al. (2011),

$$sMIR_{DM} = 0.0336(1+0.91z)\left(\frac{M_{halo}}{10^{12} M_\odot}\right)^{0.06} \sqrt{\Omega_m(1+z)^3 + \Omega_\Lambda} \quad Gyr^{-1} \quad (2)$$

As discussed in Dekel et al. (2013) the weak dependence of the $sMIR_{DM}$ on the halo mass (i.e. the power of 0.06) reflects the logarithmic slope of the fluctuation power spectrum and for simplicity we ignore this weak mass dependence in our analysis. Other analytic forms of the average $sMIR_{DM}$, such as those used in Lilly et al. (2013) and Dekel et al. (2013), have produced very similar results.

(b) *Gas accretion of the galaxy*

We assume that the baryonic accretion is regulated by the halo assembly and we also assume that all the accreted baryonic matter is in the form of pristine gas without stars. The average gas inflow rate of the galaxy $\Phi$ is assumed to scale with the DM halo growth rate as

$$\Phi = f_{gal} f_b \frac{dM_{halo}}{dt} \quad (3)$$

where $f_b = 0.155$ is cosmic baryon fraction ($\Omega_b / \Omega_m$) determined from Planck Collaboration et al. (2013). As in Lilly et al. (2013), $f_{gal}$ is the fraction of incoming baryons that flow from the surroundings into the halo and then penetrate down to enter the galaxy as baryonic gas. $f_{gal}$ is equivalent to the accretion efficiency $\epsilon_{in}$ in Bouché et al. (2010), the preventive feedback parameter $\zeta$ in Davé et al. (2012) and the penetration parameter $p$ in Dekel et al. (2013 & 2014). The value of $f_{gal}$ may depend on $M_{halo}$, feedback and epoch. The typical value of $f_{gal}$ is deduced to be of order ~ 0.5 from hydro-cosmological simulations (Dekel et al. 2013), which is appropriate for a crude comparison with observation (Dekel et al. 2014). Since we use the actual gas inflow rate of the galaxy $\Phi$ as the input parameter of the gas regulator model, the dynamics of the model presented in this work is largely independent of the adopted value of $f_{gal}$. However, when comparing model predictions to observations (in our future work), such as the stellar-to-halo mass (SHM) ratio, the value of $f_{gal}$ does matter.

(c) *Star formation and outflow*

As a consequence of the Schmidt-Kennicutt relation (Kennicutt 1998), the instantaneous average SFR of the galaxy is closely related to the gas mass present within the galaxy and we thus linked these two quantities together via the star-formation efficiency ε, as

$$\text{SFR} = \varepsilon \, M_{gas} \quad (4)$$

In fact Equation (4) can be regarded as the definition of ε. Then the gas depletion timescale $\tau_{dep}$ is given by $\tau_{dep} = M_{gas}/\text{SFR} = 1/\varepsilon$. It should be noted that in our analysis, $M_{gas}$ always refers to the total gas mass within the galaxy and it includes both atomic and molecular gas. Although the stars are actually being formed only out of the molecular gas, we do not differentiate between the atomic and molecular gas in this paper for simplicity. It should also be noted that Equation (4) does not necessarily imply a Schmidt-Kennicutt relation with a power law index N=1, since the star-formation efficiency ε is expected to scale with the stellar mass (or halo mass) and thus to scale with the gas mass as well. The effective power law index N from Equation (4) hence also depends on the exact analytic from of ε and its dependence on $M_{gas}$.

The mass-loss rate of the galaxy Ψ, i.e. the outflow, is very likely to be closely related to the average SFR of the galaxy. Analogous to Equation (4), we link these two quantities together via λ, as

$$\Psi = \lambda \cdot SFR \quad (5)$$

where λ is the mass-loading factor. Similar to ε, Equation (5) can be regarded as the definition of λ.

Star-formation efficiency ε and mass-loading factor λ are two central parameters in modeling galaxy formation and evolution. Both ε and λ are expected to scale with stellar mass (or halo mass). For a given galaxy, since the stellar mass of the galaxy will increase with time via star formation, the values of ε and λ are also expected to evolve with time. A full treatment of ε and λ will be presented in our future work, as the analytic form of ε and λ can only be determined from the combination of different observational constraints to break the degeneracy between ε and λ (see the discussion in Section 2.3). In this paper, we first assume both ε and λ to be constant over time for a given galaxy and then study how different values of ε and λ will change the results. Since in this paper we mainly focus on studying the dynamics of the gas regulator model rather than comparing the model predictions to observations, the exact values of ε and λ are less relevant. Although in Section 4 we do compare the predicted sSFR-history to the observed values, we will show in that section that the predicted sSFR-history in the gas regulator model is largely insensitive to the values of ε and λ.

(d) *The evolution of gas mass*

It is straightforward to see from the mass conservation that the change of the total baryonic mass of the galaxy, $M_B$, (defined as $M_B = M_{gas} + M_{star}$) per unit time is

$$\frac{dM_B}{dt} = \Phi - \Psi \quad (6)$$

The change of stellar mass of the galaxy per unit time is

$$\frac{dM_{star}}{dt} = (1-R) \cdot SFR \quad (7)$$

where R is the fraction of the mass of the newly formed stars as measured by the SFR, which is quickly returned to the interstellar medium (ISM), through stellar winds and supernovae. We will assume instantaneously in practice, with the remaining (1−R) staying in the form of long-lived stars. Thus (1−R)SFR is the net SFR that contributes to the net stellar mass increase of the galaxy.

Putting Equations (6) and (7) together, the change of the gas mass of the galaxy per unit time is given by

$$\begin{aligned}\frac{dM_{gas}}{dt} &= \Phi - (1-R)\cdot SFR - \Psi \\ &= \Phi - (1-R+\lambda)\varepsilon M_{gas}\end{aligned} \quad (8)$$

(e) *Metal production*

There are two sources of metals for a given galaxy. The main source of the metals is from star formation. The total metal mass produced by star formation per stellar generation is $y$*SFR, where $y$ is the average yield per stellar generation and $y$ is assumed to be a constant, independent of both epoch and stellar mass. The second source of the metals is from enriched inflows and in this case the metal supply rate is $Z_0\Phi$, where $Z_0$ is the metallicity of the infalling gas.

There are three destinations of the produced metals. The part of the metals that is locked up into long-lived stars is $Z_{gas}$(1-R)*SFR, where $Z_{gas}$ is the metallicity of the gas and is defined as $Z_{gas} = M_{Z,gas} / M_{gas}$ and $M_{Z,gas}$ is the mass of metals in the gas reservoir. The part of the metals that is expelled from the galaxy as outflow is $Z_{gas}\Psi$. The rest of the metals that is added to the gas reservoir of the galaxy is $dM_{Z,gas}/dt$. From the mass conservation of the metals, it is straightforward to write

$$y\cdot SFR + Z_0\Phi = Z_{gas}(1-R)\cdot SFR + Z_{gas}\Psi + \frac{dM_{Z,gas}}{dt} \quad (9)$$

(f) *Summary*

The gas regulator model presented in different literature may be slightly different in the detailed implementations or with different assumptions. For instance, in Davé et al. (2012) the gas mass is assumed to be constant with epoch, as they find that star-forming galaxies in hydrodynamic simulations are usually seen to lie near the equilibrium condition; in Dayal et al. (2013) the gas inflow rate is assumed to be proportional to the SFR in order to obtain the simplest analytical solution. The five equations - (4)(5)(7)(8) and (9) are the most common basic equations of the typical gas regulator model. Equation (4) is the star formation law and Equation (5) links the wind outflow to the star formation rate. Equation (4) and (5) can also be regarded as the definition of the star formation efficiency and mass-loading factor. Equation (7)(8)(9) describe the change in stellar mass, gas mass and metal mass respectively.

**2.2 Analytic Solutions**

In this section we derive the general analytic forms of the evolution of the key galaxy properties, such as the gas mass, SFR, stellar mass, sSFR and metallicity in the gas regulator model, in terms of the basic input parameters i.e. $\Phi$, $\varepsilon$, $\lambda$, $\tau_{eq}$ and R. The results are summarized in Table 1. In particular, unlike many similar studies, we do not generally assume the galaxy to live around the equilibrium state. As it will be discussed later in Section 5, gas rich low mass galaxies are very unlikely to live around the equilibrium state at any epoch.

Before discussing the results, we wish to clarify that all results presented in the current paper are applied to star-forming galaxies only. Quenching processes can be incorporated into the model by modifying Φ, λ or ε. For instance, reduce the gas inflow Φ to represent some external mechanism that cuts off the cold gas inflow of the galaxy such as strangulation (Larson et al. 1980; Balogh et al. 2000; Balogh & Morris 2000) or halo mass quenching (e.g. Birnboim & Dekel 2003; Kereš et al. 2005); reduce the star-formation efficiency ε to represent some internal mechanism that heats up the gas such as the radio-mode feedback (Croton et al. 2006) or due to the morphological or gravitational quenching mechanism (Martig et al. 2009 & 2013; Genzel et al. 2013); increase the mass-loading factor λ to a very large number to represent some feedback mechanism (e.g. AGN and/or stellar feedback) that can quickly expel the gas out of the galaxy and deplete the gas reservoir (e.g. Maiolino et al. 2012; Förster Schreiber et al. 2013; Cicone et al. 2014). We will incorporate quenching processes and extend the gas regulator model to the passive galaxies in our future work.

**(a)** $M_{gas}(t)$

We first assume Φ, R, λ and ε are all constant or only change slowly with time, Equation (8) can be solved analytically in a simple way and the gas mass as a function of time is given by

$$M_{gas}(t) = [M_{gas}(t_0) - \frac{\Phi}{\varepsilon(1-R+\lambda)}]e^{-\varepsilon(1-R+\lambda)(t-t_0)} + \frac{\Phi}{\varepsilon(1-R+\lambda)} \qquad (10)$$

where $t$ is the Hubble time and $M_{gas}(t_0)$ is the initial gas mass at some earlier time of $t_0$. In Section 3.1 we will show that with realistically evolving gas inflow Φ determined from cosmological simulation, the analytic solution given by Equation (10) is a very good approximation to the exact numeric solution.

Equation (10) clearly shows the dynamical behavior of the gas mass evolution of the galaxy. The equilibrium gas mass $M_{gas,eq}$ is simply given by the last term in Equation (10) as

$$M_{gas,eq} = \frac{\Phi}{\varepsilon(1-R+\lambda)} \qquad (11)$$

The term of $-\varepsilon(1-R+\lambda)$ in the exponent in Equation (10) determines how fast the galaxy gas mass can reach the equilibrium gas mass and the equilibrium timescale is given by

$$\tau_{eq} = \frac{\tau_{dep}}{1-R+\lambda}$$
$$= \frac{1}{\varepsilon(1-R+\lambda)} \qquad (12)$$

where $\tau_{dep} = 1/\varepsilon$ is the gas depletion time. We will see later that $\tau_{eq}$ is the central parameter in the gas regulator model that is essentially in control of the evolution of all key galaxy properties. The dynamical behavior of the gas mass evolution becomes clearer if we reformulate Equation (10) in terms of $M_{gas,eq}$ and $\tau_{eq}$

$$M_{gas}(t) = [M_{gas}(t_0) - M_{gas,eq}]e^{-\frac{\Delta t}{\tau_{eq}}} + M_{gas,eq} \qquad (13)$$

where $\Delta t$ is the time interval between $t_0$ and $t$, i.e $\Delta t = t - t_0$. The term in the square bracket tells how far the initial gas mass is away from the equilibrium gas mass. The impact of this term, no matter it is negative (i.e. the initial gas mass is less than the equilibrium gas mass) or positive (i.e. the initial gas mass is larger than the equilibrium gas mass), decreases with time. It becomes negligible when $\Delta t \gg \tau_{eq}$ and the gas mass then reaches the equilibrium gas mass.

If we set $t_0 \sim 0$ (then $\Delta t = t - t_0 = t$) and assume $M_{star}(t_0) \sim 0$ in Equation (13), Equation (13) turns into

$$M_{gas}(t) = M_{gas,eq}(1 - e^{-\frac{t}{\tau_{eq}}})$$
$$= \Phi \tau_{eq}(1 - e^{-\frac{t}{\tau_{eq}}}) \quad (14)$$

**(b) SFR(t)**

By multiplying ε on both sides of Equation (10), the SFR as a function of time is given by

$$SFR(t) = [SFR(t_0) - \frac{\Phi}{(1-R+\lambda)}]e^{-\varepsilon(1-R+\lambda)(t-t_0)} + \frac{\Phi}{1-R+\lambda} \quad (15)$$

where SFR($t_0$) is the initial SFR at some earlier time of $t_0$. The SFR in the equilibrium state, i.e. at $M_{gas,eq}$, is given by

$$SFR_{eq} = \frac{\Phi}{1-R+\lambda} \quad (16)$$

The timescale for the SFR to reach the equilibrium SFR is given by $\tau_{eq}$. Similar to Equation (13), we can reformulate Equation (15) in terms of $SFR_{eq}$ and $\tau_{eq}$

$$SFR(t) = [SFR(t_0) - SFR_{eq}]e^{-\frac{\Delta t}{\tau_{eq}}} + SFR_{eq} \quad (17)$$

If we set $t_0 \sim 0$ (then $\Delta t = t - t_0 = t$) and assume SFR($t_0$) ~ 0 in Equation (17), Equation (17) turns into

$$SFR(t) = SFR_{eq} \cdot (1 - e^{-\frac{t}{\tau_{eq}}})$$
$$= \Phi \tau_{eq} \varepsilon (1 - e^{-\frac{t}{\tau_{eq}}}) \quad (18)$$

Alternatively, Equation (18) can be simply obtained by multiplying ε on both sides of Equation (14).

**(c) $M_{star}(t)$**

If the stellar mass is defined as the *actual* stellar mass of surviving long-lived stars in the galaxy, noted as $M_{star}$, then $M_{star}(t)$ is given by integrating the reduced SFR history, i.e. by integrating (1-R)*SFR(t), as

$$M_{star}(t) = M_{star}(t_0) + \int_{t_0}^{t}(1-R) \cdot SFR(t)\,dt$$
$$= M_{star}(t_0) + \tau_{eq}(1-R)[SFR(t_0) - SFR_{eq}][1 - e^{-\frac{\Delta t}{\tau_{eq}}}] + SFR_{eq} \cdot (1-R)\Delta t \quad (19)$$

where $M_{star}(t_0)$ is the initial stellar mass at some earlier time of $t_0$ and this can be seen by letting $\Delta t = 0$ in Equation (19). From Equation (19), it is clear that the stellar mass, for the first-order, is given by the product of the equilibrium SFR and the time interval $\Delta t$ (i.e. the last term on the RHS). This first-order result is then modified by the middle term, which corrects the effect of any mass difference due to the deviation of the initial SFR from the equilibrium SFR.

If we set $t_0 \sim 0$ (then $\Delta t = t - t_0 = t$) and assume $M_{star}(t_0) \sim 0$, SFR($t_0$) ~ 0 in Equation (19), Equation (19) turns into

$$M_{star}(t) = SFR_{eq} \cdot (1-R)t - \tau_{eq} SFR_{eq} \cdot (1-R)[1 - e^{-\frac{t}{\tau_{eq}}}]$$
$$= \Phi \tau_{eq} \varepsilon (1-R)[t - \tau_{eq}(1 - e^{-\frac{t}{\tau_{eq}}})] \quad (20)$$

Alternatively, the stellar mass can be defined as the integral of the SFR history and we note the stellar mass under this definition as $M_{star,int}$, to be distinguished from the actual stellar mass $M_{star}$. The difference between these two definitions of stellar mass is simply a factor of (1-R), i.e. $M_{star}(t) = (1-R)*M_{star,int}(t)$. Following Equation(20), $M_{star,int}(t)$ is given by

$$M_{star,int}(t) = \Phi \tau_{eq} \varepsilon [t - \tau_{eq}(1 - e^{-\frac{t}{\tau_{eq}}})] \qquad (21)$$

**(d) sSFR(*t*)**

If the sSFR is defined as the SFR over the actual stellar mass of the galaxy, i.e. $SFR/M_{star}$, then the sSFR as a function of time is given by putting Equation (17) and (19) together,

$$sSFR(t) = \frac{[SFR(t_0) - SFR_{eq}]e^{-\frac{\Delta t}{\tau_{eq}}} + SFR_{eq}}{M_{star}(t_0) + \tau_{eq}(1-R)[SFR(t_0) - SFR_{eq}][1 - e^{-\frac{\Delta t}{\tau_{eq}}}] + SFR_{eq} \cdot (1-R)\Delta t} \qquad (22)$$

Now, we can set $t_0 \sim 0$ (then $\Delta t = t - t_0 = t$) and assume $M_{star}(t_0) \sim 0$, $SFR(t_0) \sim 0$ in Equation (22), and Equation (22) turns into

$$sSFR(t) = \frac{1}{1-R} \frac{1 - e^{-\frac{t}{\tau_{eq}}}}{t - \tau_{eq}[1 - e^{-\frac{t}{\tau_{eq}}}]} \qquad (23)$$

The equilibrium value of the sSFR is reached when $t >> \tau_{eq}$ and is given by

$$sSFR_{eq} = \frac{1}{1-R} \frac{1}{t - \tau_{eq}} \qquad (24)$$
$$\sim \frac{1}{1-R} \frac{1}{t}$$

The sSFR can also be defined as the SFR over the integrated stellar mass of the galaxy, i.e. $SFR/M_{star,int}$, and we note the sSFR under this definition as $sSFR_{int}$. It is clear that $sSFR_{int}(t) = (1-R)*sSFR(t)$ and $sSFR_{int}(t)$ is given by

$$sSFR_{int}(t) = \frac{1 - e^{-\frac{t}{\tau_{eq}}}}{t - \tau_{eq}[1 - e^{-\frac{t}{\tau_{eq}}}]} \qquad (25)$$

The equilibrium value of the $sSFR_{int}$ when $t >> \tau_{eq}$ is given by

$$sSFR_{int,eq} = \frac{1}{t - \tau_{eq}} \qquad (26)$$
$$\sim \frac{1}{t}$$

**(e) $f_{gas}(t)$**

Following the usual definition of the gas fraction, i.e. $f_{gas} = M_{gas} / (M_{star} + M_{gas})$, putting together Equation (14) and (20) gives

$$f_{gas}(t) = \cfrac{1}{1+\varepsilon(1-R)\left(\cfrac{t}{1-e^{-\frac{t}{\tau_{eq}}}}-\tau_{eq}\right)} \quad (27)$$

When $t \ll \tau_{eq}$, $e^{-t/\tau_{eq}} \sim 1-t/\tau_{eq}$, then $f_{gas}=1$. When $t \gg \tau_{eq}$, the equilibrium gas fraction is given by

$$\begin{aligned} f_{gas,eq}(t) &= \frac{1}{1+\varepsilon(1-R)(t-\tau_{eq})} \\ &\sim \frac{1}{1+\varepsilon(1-R)t} \end{aligned} \quad (28)$$

**(f) $\Psi(t)$**

The outflow as a function of time is given by simply inserting Equation (18) into Equation (5)

$$\begin{aligned} \Psi(t) &= \Phi\, \tau_{eq}\varepsilon\lambda(1-e^{-\frac{t}{\tau_{eq}}}) \\ &= \Phi \frac{\lambda}{1-R+\lambda}(1-e^{-\frac{t}{\tau_{eq}}}) \end{aligned} \quad (29)$$

When $t \gg \tau_{eq}$, the equilibrium outflow is given by

$$\Psi_{eq}(t) = \Phi \frac{\lambda}{1-R+\lambda} \quad (30)$$

**(g) $Z_{gas}(t)$**

As in Section 2.1 (e), following the definition of gas phase metallicity $Z_{gas}$, the change of $Z_{gas}$ per unit time is given by

$$\frac{dZ_{gas}}{dt} = \frac{1}{M_{gas}}\frac{dM_{Z,gas}}{dt} - \frac{Z_{gas}}{M_{gas}}\frac{dM_{gas}}{dt} \quad (31)$$

By reformulating Equation (9) to obtain $dM_{Z,gas}/dt$ and inserting it with Equation (8) into (31), it gives

$$\frac{dZ_{gas}}{dt} = y\varepsilon - (Z_{gas}-Z_0)\frac{\Phi}{M_{gas}} \quad (32)$$

Similar to Equation (10), if we assume $y$, $\varepsilon$, $\Phi$, $M_{gas}$ are all constant or only change slowly with time, Equation (32) can be solved analytically in a simple manner and the gas phase metallicity as a function of time is given by

$$Z_{gas}(t) = Z_0 + y\frac{SFR}{\Phi} + [Z_{gas}(t_0)-Z_0-y\frac{SFR}{\Phi}]e^{-\frac{\Phi}{M_{gas}}(t-t_0)} \quad (33)$$

$Z_{gas}(t_0)$ is the initial gas phase metallicity at some earlier time of $t_0$. Again, we will show in Section 3.1 that with realistically evolving $\Phi$ and $M_{gas}$, the analytic solution given by Equation (33) is a very good approximation to the exact numeric solution.

Set $t_0 \sim 0$ and assume $Z_{gas}(t_0) \sim 0$, and Equation (33) turns into

$$Z_{gas}(t) = [Z_0 + y\frac{SFR}{\Phi}][1 - e^{-\frac{\Phi}{M_{gas}}t}] \quad (34)$$

Insert Equation (14)(18) into (34) finally gives

$$Z_{gas}(t) = [Z_0 + y\tau_{eq}\varepsilon(1 - e^{-\frac{t}{\tau_{eq}}})][1 - e^{-\frac{t}{\tau_{eq}(1-e^{-t/\tau_{eq}})}}] \quad (35)$$

It should be noted that many authors have shown that the equilibrium gas metallicity is given by $y*SFR/\Phi$ if the infalling gas is about pristine, i.e. $Z_0=0$ (e.g. Finlator et al. 2008; Lilly et al. 2013) and the timescale for a galaxy to return to this equilibrium gas metallicity after a metallicity-perturbing interaction is given by the dilution time $M_{gas}/\Phi$ as argued in Finlator et al. (2008) (our $\Phi$ is equivalent to the $\dot{M}_{ACC}$ in Finlator et al. 2008). This seems to be correct from Equation (34), as $M_{gas}/\Phi$ appears explicitly in the time decay term in the second square bracket. However, it becomes clear from Equation (35) that the true equilibrium timescale of the gas metallicity is controlled by both $\tau_{eq}$ and $M_{gas}/\Phi$ (which is equal to $\tau_{eq}(1-e^{-t/\tau_{eq}})$), and depends on which timescale is longer. Apparently $\tau_{eq}$ is always larger than or equal to $M_{gas}/\Phi = \tau_{eq}(1-e^{-t/\tau_{eq}})$, therefore the equilibrium timescale of the gas metallicity is primarily controlled by $\tau_{eq}$. When $t \gg \tau_{eq}$, the equilibrium gas phase metallicity is given by

$$\begin{aligned} Z_{gas,eq} &= Z_0 + y\tau_{eq}\varepsilon \\ &= Z_0 + \frac{y}{1-R+\lambda} \end{aligned} \quad (36)$$

In fact, when $t \ll \tau_{eq}$, $e^{-t/\tau_{eq}} \sim 1-t/\tau_{eq}$, then the term in the second square bracket in Equation (35) has a lower limit of

$$1 - e^{-\frac{t}{\tau_{eq}(1-e^{-t/\tau_{eq}})}}\bigg|_{t \ll \tau_{eq}} \sim 1 - e^{-1} \sim 0.63 \quad (37)$$

When $t \gg \tau_{eq}$, the value of this term goes to unity. Therefore, when $t > 0$ the term in the second square bracket in Equation (34) or (35) has a value ranging from 0.63 to unity, i.e. this term can only change $Z_{gas}$ by a factor less than two at most. In other words, the metallicity perturbation controlled by the dilution time of $M_{gas}/\Phi$ can only change the gas metallicity by a factor of less than two at most. Therefore, the "equilibrium" gas metallicity of $Z_0+y*SFR/\Phi$ as in Finlator et al. (2008) and Lilly et al. (2013) is a good approximation to the more general solution as given by Equation (34) or (35).

The time-dependent evolution of $Z_{gas}$ is primarily controlled by the $(1-e^{-t/\tau_{eq}})$ term in the first square bracket in Equation (35), which has a value ranging from zero to unity. The scatter of the mass-metallicity relation is hence primarily governed by $\tau_{eq}$, not by the dilution time of $M_{gas}/\Phi$. In fact, it is clear from Table 1 that the scatters in most of the key scaling relations such as the $M_{star}$-SFR relation, $M_{star}$-$Z_{gas}$ relation, $M_{star}$-$f_{gas}$ relation are all primarily governed by $\tau_{eq}$.

**(h) $Z_{star}(t)$**
Similar to the gas metallicity, we define $M_{z,star}$ as the mass of metals locked in the long-lived stars and $Z_{star}$ as the stellar metallicity, which is given by $Z_{star} = M_{z,star}/M_{star}$. It should be noted that the stellar metallicity derived here is the *mass averaged* stellar metallicity over all the long-lived stars in the galaxy. Individual stars in the same galaxy may have very different metallicities. The old stars are expected to form from less metal enriched gas and thus have lower metallicities, while the newly formed stars are expected to have higher metallicities. Following the definition of $Z_{star}$, the change of $Z_{star}$ is given by

$$\frac{dZ_{star}}{dt} = \frac{1}{M_{star}} \frac{dM_{Z,star}}{dt} - \frac{Z_{star}}{M_{star}} \frac{dM_{star}}{dt} \quad (38)$$

The change rate of the metal locked up into long-lived stars is given by

$$\frac{dM_{Z,star}}{dt} = Z_{gas}(1-R) \cdot SFR \quad (39)$$

Inserting Equation (7) and (39) into (38) gives

$$\frac{dZ_{star}}{dt} = sSFR \cdot (1-R)(Z_{gas} - Z_{star}) \quad (40)$$

If we assume both sSFR and $Z_{gas}$ to be constant or only change slowly with time, $Z_{star}$ as a function of time is given by

$$Z_{star}(t) = Z_{gas} + [Z_{star}(t_0) - Z_{gas}]e^{-sSFR \cdot (1-R)(t-t_0)} \quad (41)$$

Set $t_0 \sim 0$ and assume $Z_{star}(t_0) \sim 0$, and Equation (41) turns into

$$Z_{star}(t) = Z_{gas}[1 - e^{-sSFR \cdot (1-R)t}] \quad (42)$$

For the lowest order, the sSFR can be approximated by $1/[(1-R)t]$ as in Equation (24). Substitute it into Equation (42) and it gives

$$Z_{star}(t) \sim Z_{gas}(1 - e^{-1}) \\ \sim 0.63 Z_{gas} \quad (43)$$

This suggests that for a given galaxy, or for galaxies with similar stellar mass, the stellar metallicity is about ~0.2 dex lower than the gas metallicity, in broad agreement with observations (e.g. Halliday et al. 2008) and predictions from more realistic cosmological simulations/models (e.g. Finlator et al. 2008; Pipino et al. 2014).

From Equation (42), the equilibrium stellar metallicity is simply

$$Z_{star,eq} = Z_{gas} \quad (44)$$

The equilibrium timescale for the stellar metallicity to catch up the gas metallicity is $1/[sSFR(1-R)]$ if the gas metallicity $Z_{gas}$ is constant. In reality, $Z_{gas}$ is expected to evolve with epoch and therefore the actual equilibrium timescale of the stellar metallicity also depends on the detailed evolution history of $Z_{gas}$, i.e. also depends on the equilibrium timescale $\tau_{eq}$. Again, although the simple analytical solution given in Equations (41) and (42) are derived under the assumption that both sSFR and $Z_{gas}$ are constant or only change slowly with time, in Section 3.1 we will show that with realistically evolving sSFR and $Z_{gas}$, this simple analytical solution is a very good approximation to the exact numeric solution.

**(i)  $f_{star}(t)$, $f_{out}(t)$ and $f_{res}(t)$**

As in Lilly et al. (2013), it is useful to derive the fractional distribution of the inflowing gas as $f_{star}$, $f_{out}$ and $f_{res}$, which describe the three destinations of the incoming baryons: locked in stars, re-expelled in outflows, and added to the gas reservoir. With the use of $\tau_{eq}$, we can greatly simplify the results in Lilly et al. (2013). The fraction of the inflow that is locked up into long-lived stars is given by

$$f_{star}(t) = \frac{(1-R) \cdot \text{SFR}}{\Phi} \qquad (45)$$
$$= (1-R)\tau_{eq}\varepsilon(1-e^{-\frac{t}{\tau_{eq}}})$$

The fraction of the inflow that is expelled from the galaxy as outflow is given by

$$f_{out}(t) = \frac{\Psi}{\Phi} \qquad (46)$$
$$= \tau_{eq}\varepsilon\lambda(1-e^{-\frac{t}{\tau_{eq}}})$$

The fraction of the inflow that is added to the gas reservoir of the galaxy is given by

$$f_{res}(t) = \frac{dM_{gas}/dt}{\Phi} \qquad (47)$$
$$= e^{-\frac{t}{\tau_{eq}}}$$

Clearly, $f_{star} + f_{out} + f_{res} = 1$ at any epoch. At $t=0$, $f_{res}=1$, i.e. all the inflow is added to the gas reservoir. In the equilibrium state at $t \gg \tau_{eq}$, $f_{res}=0$, i.e. the galaxy has reached a balance between the inflow, outflow and SFR, and the gas mass remains constant with time.

**(j) Timescales: $\tau_{dep}$, $\tau_{eq}$ and $\tau_{dil}$**

In our analysis above, in addition to the Hubble timescale which is the local age of the universe, there are three involved important and closely related timescales: gas depletion timescale $\tau_{dep}$, equilibrium timescale $\tau_{eq}$ and dilution timescale $\tau_{dil}$. We have showed above that $\tau_{eq}$ is the equilibrium timescale that governs the evolution of the galaxy population, since $\tau_{eq}$ appears explicitly in the time decay term in essentially all the key galaxy properties (see the summary in Table 1). The equilibrium timescale, by definition, is the timescale for a galaxy to return to its equilibrium state from a perturbation or from an (arbitrary) initial condition, therefore the scatters in most of the key scaling relations, such as the $M_{star}$-SFR relation, $M_{star}$-$Z_{gas}$ relation, $M_{star}$-$f_{gas}$ relation, are all primarily governed by $\tau_{eq}$, not by $\tau_{dep}$ and $\tau_{dil}$.

The gas depletion timescale $\tau_{dep}$ is equal to $1/\varepsilon$ and hence it is primarily determined by the star formation law. $\tau_{dep}$ is closely related to $\tau_{eq}$ via Equation (12). If the value of the mass-loading factor of the galaxy is around the mass return fraction R, $\tau_{dep}$ is comparable to $\tau_{eq}$. A larger mass-loading factor will make $\tau_{eq}$ shorter than $\tau_{dep}$, which means it will be faster for the galaxy to reach the equilibrium state, or re-equilibrate itself quicker from any perturbation that drives the galaxy away from the equilibrium.

The dilution timescale $\tau_{dil}$, as introduced in Finlator et al. (2008) and Davé et al. (2012), is defined as $\tau_{dil} = M_{gas}/\Phi$. Inserting Equation (14), it gives

$$\tau_{dil} = \tau_{eq}(1 - e^{-\frac{t}{\tau_{eq}}}) \qquad (48)$$

Apparently, $\tau_{dil}$ is always shorter than or by maximum equal to $\tau_{eq}$. In particular, when $t \ll \tau_{eq}$ (i.e. out-of-equilibrium state), $e^{-t/\tau_{eq}} \sim 1 - t/\tau_{eq}$, then $\tau_{dil} \sim t$. When $t \gg \tau_{eq}$ (i.e. in equilibrium state), $\tau_{dil} \sim \tau_{eq}$. It should be noted that the $\tau_{dil}$ given by Equation (14) in Davé et al. (2012), i.e. $\tau_{dil} = \tau_{eq}$ (their η is referred to our λ), is the equilibrium case (when $t \gg \tau_{eq}$) of the more general solution given by Equation (48) above.

Putting together and assuming R=0, it gives

$\tau_{dil} \leq \tau_{eq} \leq \tau_{dep}$ (49)

A non-zero R with a small λ may make $\tau_{eq}$ slightly larger than $\tau_{dep}$ (see Equation (12)).

It should be noted that the out-of-equilibrium case (when $t \ll \tau_{eq}$) has largely been ignored in most of the gas-regulator or bathtub model work for simplicity. These work often assumes that the galaxies are living around the equilibrium state, i.e. assume $t \gg \tau_{eq}$. We have shown above that the equilibrium timescale is uniquely determined by $\tau_{eq}$ and we will show later in Section 5 the value of $\tau_{eq}$ estimated from observations, which clearly suggests that most low mass star-forming galaxies are very likely to be out of equilibrium across most of the cosmic time. Therefore, the general solutions derived above (and summarized as the semi-exact solution in Table 1) without assuming any equilibrium condition are important improvement to complete the framework of the current gas-regulator or bathtub model. The equilibrium solutions as shown in many similar work (and summarized in Table 1) are the special cases of the general solutions at $t \gg \tau_{eq}$.

## 2.3 Dynamic Behaviors

In Table 1, we summarize the basic input parameters, the derived key galaxy properties and the fractional splitting of the inflow in the gas regulator model for star-forming galaxies. We stress that the semi-exact solutions that are derived in previous sub-section and summarized in Table 1 are *exact* analytic solutions if Φ, R, λ and ε are constant or the timescales of the change are longer than the equilibrium timescale $\tau_{eq}$. The equilibrium solutions (or steady-state solutions) listed in Table 1, which have been widely used in many similar studies, are the special cases of the semi-exact solutions, i.e. solutions in the equilibrium state at $t \gg \tau_{eq}$. In reality, all the four parameters of Φ, R, λ and ε are likely to evolve with time, especially at high redshifts (z > ~2). Therefore, the results determined from the semi-exact solutions will be different from the exact numeric solutions (that is why they are called as "semi-exact") and the difference will dependent on how fast Φ, R, λ and ε vary with time. We will show in the next section that with realistically evolving Φ determined from cosmological simulations, the semi-exact solutions summarized in Table 1 are very good approximations to the exact numeric solutions. This is because, as will be shown later in Figure 1, the gas inflow rate, halo mass and stellar mass (on which λ and ε are expected to depend) change much more slowly with time after z ~ 2. Even in some extreme situation that Φ, R, λ and ε all change rapidly with time (e.g. galaxies in some transient phase such as the starburst galaxy or galaxy in the process of being quenched), the results computed from the semi-exact solutions are largely deviated from the exact numeric solutions, we can still use these semi-exact solutions to understand the localized dynamical evolution of the galaxy properties on short timescale, i.e. we can always choose a short time interval such that within this short time interval Φ, R, λ and ε are roughly constant. We will discuss this in more details in the next section.

Table 1. Summary of the input parameters and predicted galaxy properties in the gas regulator model

*(a) Input parameters*

| Parameter | Definition | |
|---|---|---|
| gas inflow rate of the galaxy $\Phi$ | $\Phi = f_b f_{gal} \frac{dM_{halo}}{dt}$ | $f_b$ - cosmic baryon fraction<br>$f_{gal}$ - halo penetration efficiency |
| star-formation efficiency $\varepsilon$ | $\varepsilon = \text{SFR} / M_{gas}$ | |
| mass-loading factor $\lambda$ | $\lambda = \Psi / \text{SFR}$ | $\Psi$ - outflow rate |
| equilibrium timescale $\tau_{eq}$ | $\tau_{eq} = \frac{1}{\varepsilon(1 - R + \lambda)}$ | R - mass return fraction from stars |

*(b) Predicted galaxy properties*

| Galaxy Property | Semi-Exact Solution [1] | Equilibrium Solution ($t \gg \tau_{eq}$) [1] |
|---|---|---|
| $M_{gas}(t)$ | $\Phi \tau_{eq}(1 - e^{-\frac{t}{\tau_{eq}}})$ | $\Phi \tau_{eq}$ |
| SFR($t$) | $\Phi \tau_{eq} \varepsilon (1 - e^{-\frac{t}{\tau_{eq}}})$ | $\Phi \tau_{eq} \varepsilon = \frac{\Phi}{1 - R + \lambda}$ |
| $M_{star,int}(t)$ | $\Phi \tau_{eq} \varepsilon [t - \tau_{eq}(1 - e^{-\frac{t}{\tau_{eq}}})]$ | $\Phi \tau_{eq} \varepsilon (t - \tau_{eq}) \sim \Phi \tau_{eq} \varepsilon t$ |
| $M_{star}(t)$ | $(1-R) * M_{star,int}(t)$ | $(1-R) * M_{star,int,eq}(t)$ |
| sSFR$_{int}(t)$ | $\frac{1 - e^{-\frac{t}{\tau_{eq}}}}{t - \tau_{eq}(1 - e^{-\frac{t}{\tau_{eq}}})}$ | $\frac{1}{t - \tau_{eq}} \sim \frac{1}{t}$ |
| sSFR($t$) | sSFR$_{int}(t) / (1-R)$ | sSFR$_{int,eq}(t) / (1-R)$ |
| $f_{gas}(t)$ | $\frac{1}{1 + \varepsilon(1-R)\left(\frac{t}{1 - e^{-\frac{t}{\tau_{eq}}}} - \tau_{eq}\right)}$ | $\frac{1}{1 + \varepsilon(1-R)(t - \tau_{eq})} \sim \frac{1}{1 + \varepsilon(1-R)t}$ |
| $\Psi(t)$ | $\Phi \tau_{eq} \varepsilon \lambda (1 - e^{-\frac{t}{\tau_{eq}}}) = \Phi \frac{\lambda}{1 - R + \lambda}(1 - e^{-\frac{t}{\tau_{eq}}})$ | $\Phi \tau_{eq} \varepsilon \lambda = \Phi \frac{\lambda}{1 - R + \lambda}$ |
| $Z_{gas}(t)$ | $[Z_0 + y\tau_{eq}\varepsilon(1 - e^{-\frac{t}{\tau_{eq}}})][1 - e^{-\frac{t}{\tau_{eq}(1 - e^{-t/\tau_{eq}})}}]$ | $Z_0 + y\tau_{eq}\varepsilon = Z_0 + \frac{y}{1 - R + \lambda}$ |
| $Z_{star}(t)$ | $Z_{gas}[1 - e^{-sSFR \cdot (1-R)t}]$ | $Z_{gas}$ [2] |

(c) *The fractional splitting of the inflow ($f_{star} + f_{out} + f_{res} = 1$)*

| Fraction | Semi-Exact Solution [1] | Equilibrium Solution ($t \gg \tau_{eq}$) [1] |
|---|---|---|
| $f_{star}(t) = \dfrac{(1-R) \cdot \text{SFR}}{\Phi}$ | $(1-R)\tau_{eq}\varepsilon(1 - e^{-\frac{t}{\tau_{eq}}})$ | $(1-R)\tau_{eq}\varepsilon = \dfrac{1-R}{1-R+\lambda}$ |
| $f_{out}(t) = \dfrac{\Psi}{\Phi}$ | $\tau_{eq}\varepsilon\lambda(1 - e^{-\frac{t}{\tau_{eq}}})$ | $\tau_{eq}\varepsilon\lambda = \dfrac{\lambda}{1-R+\lambda}$ |
| $f_{res}(t) = \dfrac{dM_{gas}/dt}{\Phi}$ | $e^{-\frac{t}{\tau_{eq}}}$ | 0 |

(1) As in the text, the semi-exact solutions are *exact* analytic solutions for constant $\Phi$, R, $\lambda$ and $\varepsilon$, and become "semi-exact" for evolving $\Phi$, R, $\lambda$ and $\varepsilon$. The equilibrium solutions (or steady-state solutions) are the special cases of the semi-exact solutions at $t \gg \tau_{eq}$.

(2) The equilibrium timescale of $Z_{star}$ is $1/\text{sSFR}/(1-R)$ if $Z_{gas}$ is constant. As in the text, if $Z_{gas}$ evolves with epoch, the actual equilibrium timescale of $Z_{star}$ will also depend on the evolution history of $Z_{gas}$, i.e. also depend on $\tau_{eq}$.

From Table 1, we can draw the following important conclusions.

(i) For galaxies with long equilibrium timescale $\tau_{eq}$, which usually are gas rich low mass galaxies (see the discussion in Section 5), the dynamical evolution of all the key galaxy properties are critically controlled by $\tau_{eq}$ as seen from the semi-exact solutions in Table 1. For massive galaxies with short $\tau_{eq}$ (see the discussion in Section 5), these key galaxy properties are mainly controlled by the input parameters that determine the equilibrium values, shown as the equilibrium solutions in Table 1.

(ii) The gas mass evolution is driven by the gas inflow rate $\Phi$ and the equilibrium timescale $\tau_{eq}$. The star-formation efficiency $\varepsilon$ and the mass-loading factor $\lambda$ are completely equivalent in controlling the gas mass evolution, i.e. $\varepsilon$ and $\lambda$ are degenerate in controlling $M_{gas}(t)$.

(iii) The SFR evolution and stellar mass evolution are mainly driven by the gas inflow rate $\Phi$, the equilibrium timescale $\tau_{eq}$ and the star-formation efficiency $\varepsilon$, or equivalently by $\tau_{eq}$ and the mass-loading factor $\lambda$.

(iv) The sSFR$_{int}$ evolution is controlled solely by the equilibrium timescale $\tau_{eq}$, independent of the inflow rate $\Phi$ and the individual values of $\varepsilon$ and $\lambda$. Similar to $M_{gas}(t)$, $\varepsilon$ and $\lambda$ are completely equivalent in controlling the sSFR evolution, i.e. $\varepsilon$ and $\lambda$ are degenerate in controlling sSFR$_{int}(t)$. To the first order, i.e. assuming $t \gg \tau_{eq}$, sSFR$_{int}$ is simply the reverse of the Hubble timescale. The sSFR, defined by the SFR over the actual stellar mass of the galaxy, apparently also depends on R (i.e. the mass return fraction from stars). A non-zero R will increase the amplitude of the sSFR$_{int}(t)$ by a small factor of $1/(1-R)$ and maintain its logarithmic shape.

(v) The gas fraction evolution is controlled by the star-formation efficiency $\varepsilon$, equilibrium timescale $\tau_{eq}$ and mass return fraction R, but is independent of the inflow rate $\Phi$. For a given $\varepsilon$, the gas fraction drops with epoch monotonously. When $t \gg \tau_{eq}$, the gas fraction reaches its equilibrium value which is also the minimum value of $1/(1+\varepsilon(1-R)t)$. The reason why massive galaxies have low gas fraction is partly because they have shorter $\tau_{eq}$ (see the discussion in Section 5) and possibly have higher star formation efficiency.

(vi) The specific mass increase of the baryonic matter, sMIR$_B$ (defined as sMIR$_B$ = $dM_B / dt / M_B$), generally follows the evolution of sMIR$_{DM}$ (i.e. the specific mass increase rate of the haloes). If $\lambda \sim 0$, i.e. no outflow, the baryonic mass of the galaxy is then simply $f_b f_{gal} * M_{halo}$. The gas inflow rate is given by Equation (3) and sMIR$_B$ will be the same as the sMIR$_{DM}$. Since in the ideal gas regulator model, the gas inflow rate is simply regulated by the DM halo accretion rate, any outflow will act to decrease the baryonic mass of the galaxy and thus increase the sMIR$_B$.

(vii) It is clear from the analytic form of the outflow rate $\Psi$ that $\Psi$ is always smaller than or by maximum equal to the gas

inflow rate $\Phi$. Since low mass galaxies are expected to have a mass-loading factor $\lambda$ larger than unity, their $\Psi$ will evolve towards $\Phi$ and the timescale is controlled by $\tau_{eq}$. In the equilibrium state, $\Psi \sim \Phi$, i.e. little of the infalling gas has been converted into stars. This can also be seen from the equilibrium solutions of the fractional splitting of the inflow. If $\lambda >> 1$, $f_{star,eq} \sim 0$, $f_{out,eq} \sim 1$ and $f_{res,eq} \sim 0$. Massive galaxies with a small mass-loading factor will have an outflow rate smaller than the inflow rate, as they can more effectively convert the infalling gas into stars. For massive galaxies with $\lambda << 1$, the SFR will evolve towards the inflow rate $\Phi$ and in the equilibrium state, SFR $\sim \Phi$, i.e. convert all the infalling gas into stars. Likewise, from the equilibrium solutions of the fractional splitting of the inflow, if $\lambda << 1$, then $f_{star,eq} \sim 1$, $f_{out,eq} \sim 0$ and $f_{res,eq} \sim 0$.

(viii) For a given $Z_0$ and a given $y$, the evolution of the gas phase metallicity $Z_{gas}$ is controlled by the mass-loading factor $\lambda$ and $\tau_{eq}$, or equivalently by the star-formation efficiency $\varepsilon$ and $\tau_{eq}$. The maximum metallicity for all galaxies, i.e. the metallicity upper limit, is given by $Z_0 + y/(1+R)$. This maximum metallicity is reached for massive galaxies with mass-loading factor $\lambda << 1$ in the equilibrium state.

(ix) The mass averaged stellar metallicity simply evolves towards the gas metallicity and the timescale is controlled by both 1/sSFR and $\tau_{eq}$. Therefore, the exact difference between the gas metallicity and stellar metallicity depends on both 1/sSFR and $\tau_{eq}$. But for the first order, for a given galaxy or for galaxies with similar stellar mass, the stellar metallicity is about ~0.2 dex lower than the gas metallicity.

(x) Since $\tau_{eq}$ is the equilibrium timescale, i.e. the timescale for a galaxy to return to its equilibrium state from a perturbation or from an (arbitrary) initial condition, for all the key galaxy properties listed in Table 1, the scatters in most of the key scaling relations, such as the $M_{star}$-SFR relation, $M_{star}$-sSFR relation, $M_{star}$-$Z_{gas}$ relation, $M_{star}$-$f_{gas}$ relation, are all primarily governed by $\tau_{eq}$.

## 3. NUMERIC TEST

In previous section we have derived the analytic solutions of the evolution of the key galaxy properties in the gas regulator model, and discussed their dynamic behaviors. In this section, we first validate these analytic solutions by comparing them with the exact numeric solutions in Section 3.1. In Section 3.2, we study the action of $\varepsilon$, $\lambda$ and $\tau_{eq}$ in regulating the evolution of the key galaxy properties.

### 3.1 Validation of the analytic solution

First of all, we wish to clarify that in this section we do not intend to directly compare the model results with observations, but purely to test the accuracy of the analytic solutions in Table 1 by comparing them with the exact numeric solutions. Since the equilibrium solutions in Table 1 are just the special case of the more general semi-exact solutions at $t >> \tau_{eq}$ (or simply the $\tau_{eq}$ is short, e.g. $\tau_{eq} < \sim 1$ Gyr), we want to particularly test the uncertainties of the semi-exact solutions, which are for galaxies with long $\tau_{eq}$ and out of the equilibrium state. For this purpose, we intentionally choose a long $\tau_{eq} = 10$ Gyr, which is longer than the Hubble timescale at early epochs and is comparable to the Hubble timescale at late epochs. We set $\varepsilon$ to an arbitrary (constant) value of 0.05 Gyr$^{-1}$ and $\lambda$ to an arbitrary (constant) value of unity. We will show later in Section 3.2 how the results will change with different values of $\varepsilon$ and $\lambda$. The adopted values of $\tau_{eq}$ and $\varepsilon$ may be on the extreme side (see the discussion in Section 5), however, this will help us to better assess the accuracy of the semi-exact solutions and to better understand the dynamical behaviors of the model.

We initiate the evolution of a dark matter halo with an (arbitrary) initial mass of $M_{halo}=10^6$ M$_\odot$ at $t$=0.1 Gyr. The halo mass increases with time according to the sMIR$_{DM}$ given by Equation (2) and is plotted as the green curve in the top right panel of Figure 1. It reaches a final halo mass of $\sim 10^{12}\,M_\odot$ at $t = 13.7$ Gyr ($z\sim0$). The associated halo accretion rate $dM_{halo}/dt$ is plotted as the green curve in the top left panel of Figure 1. The gas inflow rate of the galaxy $\Phi$ determined from Equation (3) is plotted as the red curve in the top left panel of Figure 1. We assume the initial baryonic mass of the galaxy to be around $f_b f_{gal} * M_{halo}$ at $t=0.1$ Gyr, in the form of gas only, i.e. in this case the initial gas mass is $f_b f_{gal} * 10^6\,M_\odot$ and initial stellar mass is zero. A modest different initial condition will produce very similar results. Since, as shown in Equation (13), the impact of the initial gas mass on the gas mass at a later time decreases and becomes negligible when $t \gg \tau_{eq}$. We set R=0 for simplicity. Because for most of the key galaxy properties in Table 1, they do not have a direct dependence on R. For the galaxy properties that have a direct dependence on R (e.g. $M_{star}$ and sSFR), the effect of R can be easily added independently by multiplying or being divided by a small factor of (1-R).

The gas mass increases with time according to the gas inflow rate $\Phi$ and is plot as the red solid curve from the exact numeric solution in the top right panel of Figure 1. The red dashed curve in the same panel shows the gas mass evolution determined from the analytic semi-exact solution given by Equation (14). The SFR is simply determined by Equation (4) and is plotted as the black solid curve for the exact numeric solution in the top left panel of Figure 1. The black dashed curve in the same panel shows the SFR evolution determined from the analytic semi-exact solution given by Equation (18). The associated stellar mass evolution by integrating the SFR-history is plotted as the black solid curve for the exact numeric solution in the top right panel of Figure 1. The black dashed curve in the same panel shows the stellar mass evolution determined from the analytic semi-exact solution given by Equation (20). In the middle left panel of Figure 1, we plot the gas fraction evolution determined from the exact numeric solution in black solid curve and from the analytic semi-exact solution given by Equation (27) in black dashed curve.

In the middle right panel of Figure 1, we plot the $sMIR_{DM}$ as the green curve determined from Equation (2). The red curve shows the specific mass increase rate of the baryonic mass of the galaxy, $sMIR_B$, determined from the exact numeric solution. As discussed before, $sMIR_B$ generally follows $sMIR_{DM}$ and the difference between $sMIR_{DM}$ and $sMIR_B$ is proportional to the value of $\lambda$. If $\lambda=0$, i.e. no outflow, $sMIR_B$ will be exactly the same as $sMIR_{DM}$. The sSFR is plotted as the black solid line for the exact numeric solution and black dashed line for the analytic semi-exact solution given by Equation (23). The inverse of the Hubble timescale, $1/t_H$, is plotted as the brown curve for reference. As shown in this panel, to the first order, both of the $sMIR_{DM}$ and sSFR in the gas regulator model simply follow $1/t_H$. We will discuss the dynamical evolution of the sSFR in detail later in Section 4.

In the bottom left panel of Figure 1 we show the gas metallicity evolution determined from the exact numeric solution (black solid curve) and from the analytic semi-exact solution given by Equation (35) (black dashed curve). For comparison, we also show the equilibrium metallicity given by Equation (36) as the red solid horizontal line. Clearly, for a galaxy with a long $\tau_{eq}$ ($\tau_{eq}$ =10 Gyr in this case), the actual gas metallicity of the galaxy will evolve towards the equilibrium metallicity but may never reach it. Therefore, the actual metallicity of the galaxy is not only controlled by the mass-loading factor (which determines the equilibrium metallicity) but also critically controlled by the timescale it takes to reach the equilibrium state, i.e. by $\tau_{eq}$ (see also the discussion in Section 5). The metallicity evolution and its dependence on $\varepsilon$, $\lambda$, $\tau_{eq}$, $M_{star}$ and SFR will be explored in a separate paper.

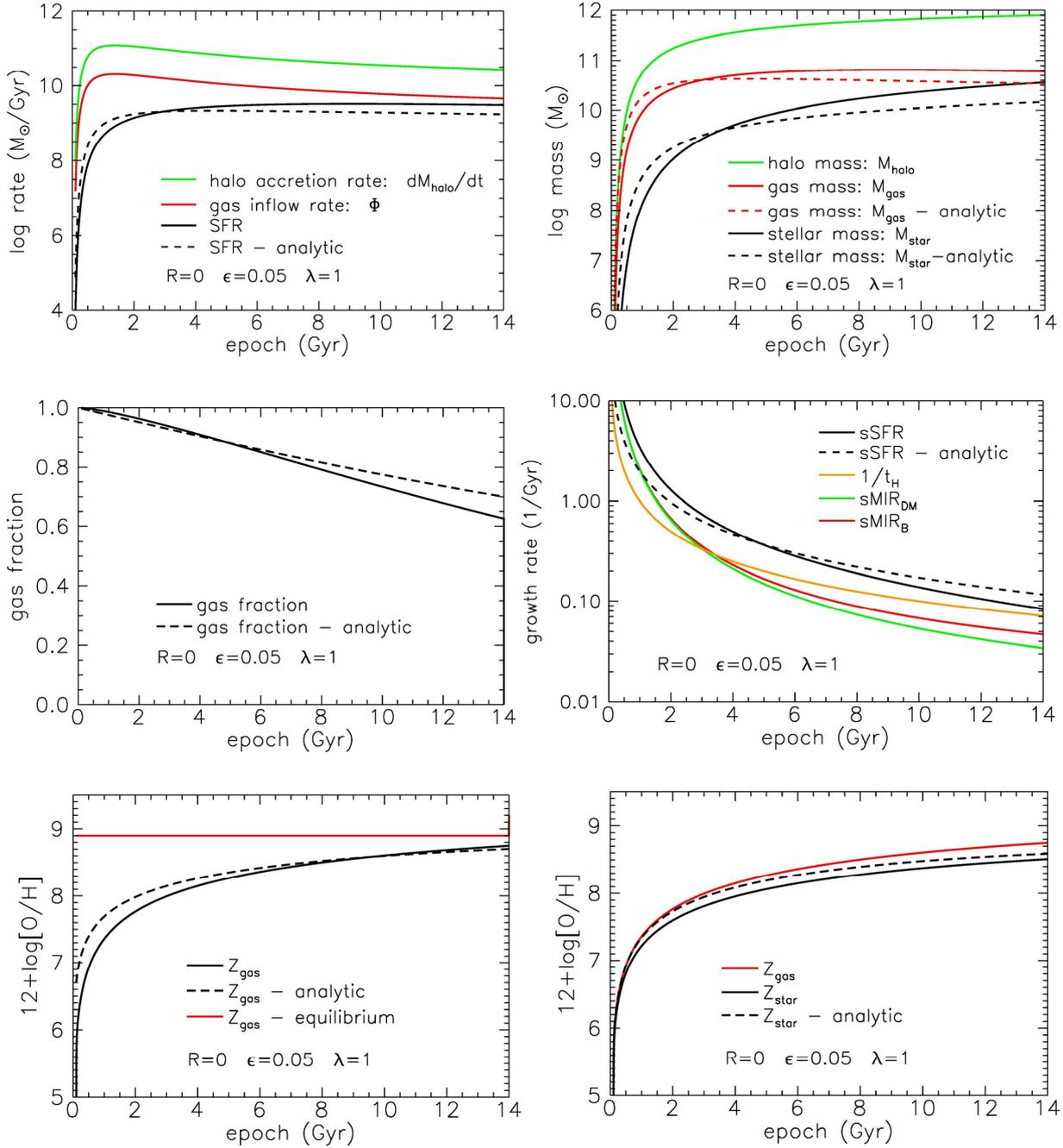

Figure 1. Evolution over cosmic time of the key galaxy properties computed from analytic semi-exact solutions (dashed lines) and exact numeric solutions (solid lines) in the gas regulator model as described in the text. It is evident that the analytic semi-exact solutions as summarized in Table 1 generally trace well the exact numeric solutions. The small differences between the analytic semi-exact solutions and the exact numeric solutions come from the fact that the gas inflow rate (red curve in the top left panel) evolves with time, especially at z>3 where the gas inflow rate increases rapidly with time.

In the bottom right panel of Figure 1, we show the mass average stellar metallicity evolution determined from the exact numeric solution (black solid curve) and from the analytic semi-exact solution given by Equation (42) (black dashed curve). It is clear that the stellar metallicity in general closely follows the evolution of the gas phase metallicity. The difference between the stellar metallicity and gas metallicity is evidently about ~0.2 dex as expected from Equation (43) but is slightly smaller at earlier epochs. This is because in this case $\tau_{eq}$ is long ($\tau_{eq}$ =10 Gyr) and hence the $Z_{gas}$ never reaches its equilibrium value (the red horizontal line in the bottom left panel of Figure 1), i.e. $Z_{gas}$ gradually increases with epoch throughout the entire cosmic time. While the equilibrium timescale for the stellar metallicity of $1/sSFR_{int}$ is shorter at earlier epochs, which means at earlier epochs it is easier (or faster) for the $Z_{star}$ to follow the evolution of $Z_{gas}$. Therefore the difference between $Z_{gas}$ and $Z_{star}$ is smaller at earlier epochs. But if $\tau_{eq}$ is very short, e.g. for massive galaxies (see discussions in Section 5), $Z_{gas}$ may have quickly reached its equilibrium value and then remains constant later on. In this case $Z_{star}$ may catch up with and then be equal to $Z_{gas}$.

As shown in Figure 1, for $M_{gas}(t)$, $SFR(t)$, $M_{star}(t)$, $sSFR(t)$, $f_{gas}(t)$, $Z_{gas}(t)$ and $Z_{star}(t)$, the analytic semi-exact solutions (which are the exact solutions for constant $\Phi$, R, $\lambda$ and $\varepsilon$ ) as summarized in Table 1 are very good approximations to the exact numeric solutions (solved with evolving $\Phi$ determined from cosmological simulations). The small differences between the analytic semi-exact solutions and the exact numeric solutions come from the fact that in reality $\Phi$ evolves with time, especially at z>3 where $\Phi$ increases rapidly with time.

### 3.2 The action of $\varepsilon$, $\lambda$ and $\tau_{eq}$ in regulating galaxy properties

Figure 1 in the previous section shows the evolution of the key galaxy properties for given (arbitrary) values of $\varepsilon$=0.05 $Gyr^{-1}$, $\lambda$=1 (i.e. $\tau_{eq}$=10 Gyr) and R=0. In Figure 2 we first plot the evolution of $M_{gas}$, SFR, $M_{star}$, $f_{gas}$ and sSFR for the same values of $\varepsilon$, $\lambda$ and R in each panel in solid lines. For the sake of clarity we show results determined from exact numeric solutions only. Then we modify the values of $\varepsilon$ and $\lambda$ (arbitrarily) by setting $\varepsilon$=0.0001 $Gyr^{-1}$, $\lambda$=1000, i.e. extreme unrealistic values, but which give the same equilibrium timescale $\tau_{eq}$=10 Gyr. Again, in this section we do not intend to make direct comparisons between the model results and observations, but purely to explore the dynamics of the model by intentionally using extreme values of $\varepsilon$ and $\lambda$, even these values are physically unrealistic in this case. This will help us to better assess the validity of the predicted dynamical behaviors of the model as seen from the analytic solutions (as discussed in Section 2.3) and to identify any potential limit of the model, i.e. any incapability of the model to reproduce observations.

We keep the same initial conditions as those used in previous section and Figure 1, i.e. an initial DM halo mass of $10^6 M_\odot$, an initial gas mass of $f_b f_{gal}*10^6 M_\odot$ and an initial stellar mass of zero at $t$=0.1 Gyr. The halo accretion rate, halo mass assembly history and gas inflow rate hence remain unchanged. The evolution of $M_{gas}$, SFR, $M_{star}$, $f_{gas}$ and sSFR for the new values of $\varepsilon$ and $\lambda$ are plotted as the dashed lines in each corresponding panel in Figure 2.

For convenience, we note the galaxy with $\varepsilon$=0.05 $Gyr^{-1}$ and $\lambda$=1 as galaxy A (solid lines in Figure 2). The other one with $\varepsilon$=0.0001 $Gyr^{-1}$ and $\lambda$=1000 is noted as galaxy B (dashed lines in Figure 2). According to Table 1 and the discussions in Section 2.3, both $SFR(t)$ and $M_{star}(t)$ depend on $\Phi$, $\varepsilon$ and $\tau_{eq}$. Galaxy A and B have the same $\Phi$ and $\tau_{eq}$, therefore it is easy to see that $SFR_A(t) / SFR_B(t) = M_{star,A}(t) / M_{star,B}(t) = \varepsilon_A / \varepsilon_B = 500$, i.e. 2.7 dex at all epochs. This is the exact difference between the black solid line and black dashed line for $SFR_A(t)$ and $SFR_B(t)$ in the top left panel of Figure 2, and for $M_{star,A}(t)$ and $M_{star,B}(t)$ in the top right panel.

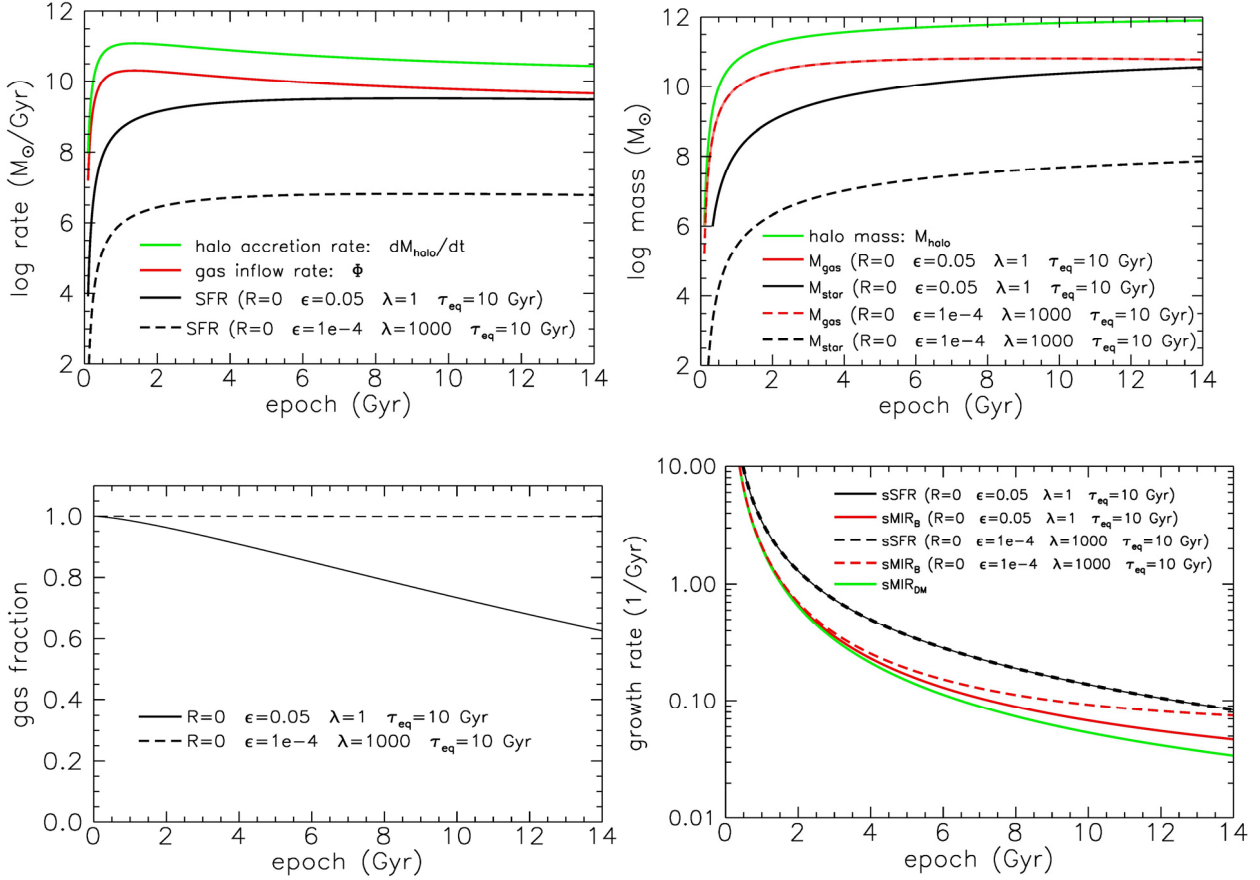

Figure 2. As Figure 1, evolution over cosmic time of the key galaxy properties computed from exact numeric solutions for two different sets of values of ε and λ, but the same $\tau_{eq}$. It should be noted that although all the results plotted here are determined from the exact numeric solutions, they still follow precisely the dynamical behaviors as seen from the analytic semi-exact solutions as described in the text. For instance, galaxies with the same $\tau_{eq}$ should have the same sSFR history. This is clearly seen in the lower right panel that the black solid line is completely overlapped with the black dashed line, although the ε and λ are very different. This has hence stressed the importance of the analytic semi-exact solutions.

For a given gas inflow rate Φ, the gas mass evolution $M_{gas}(t)$ depends only on $\tau_{eq}$, i.e. on the product of ε and λ, not on their individual values. Since galaxy A and B have the same Φ and $\tau_{eq}$, although their ε and λ are very different, they should have the same gas mass assembly history. Indeed, in the top right panel of Figure 2, the red solid line is completely overlapped with the red dashed line. It is important to note that although all the results presented in this subsection are computed from the exact numeric solutions, they still follow precisely the dynamical behaviors as predicted from the analytic semi-exact solutions, which hence emphasizes the importance of these analytic solutions.

Apart from the gas return fraction R and cosmic time $t$, the sSFR($t$) depends on only $\tau_{eq}$, which is the same for galaxy A and B. Therefore, galaxy A and B should have the same sSFR-history as predicted from the analytic semi-exact solutions. This is clearly shown in the lower right panel of Figure 2: the black solid line is completely overlapped with the black dashed line.

The gas fraction, for a given $\tau_{eq}$, is inversely proportional to ε. Therefore galaxy A has a lower gas fraction than galaxy B,

as shown in the lower left panel of Figure 2. The specific mass increase rate of the baryonic matter, sMIR$_B$, generally follows the evolution of sMIR$_{DM}$. As discussed before, outflow will act to decrease the baryonic mass of the galaxy and thus increase the sMIR$_B$. Galaxy A and B have the same $\Phi$ and $\tau_{eq}$, from Equation (29), although the mass-loading factor of galaxy B is thousand times larger than that of galaxy A, the outflow of galaxy B is only two times larger than that of galaxy A. Therefore galaxy B has a slightly larger sMIR$_B$ than that of galaxy A, as shown in the lower right panel of Figure 2.

## 4. THE ACTION OF $\tau_{eq}$ IN REGULATING sSFR AND THE DIFFICULTY TO REPRODUCE THE OBSERVED sSFR

In this section we further explore the sSFR evolution in the gas regulator model and its dependence on $\tau_{eq}$ in Section 4.1. Then in Section 4.2 we show that the predicted sSFR-history from current gas regulator model is very similar to those predicted from typical SAMs, but both are fundamentally different from the observed sSFR-history, which indicates some key process is missing in the current gas regulator model and typical SAMs.

### 4.1 The action of $\tau_{eq}$ in regulating sSFR($t$)

As discussed before, apart from the mass return fraction R and cosmic time $t$, the sSFR($t$) is solely controlled by $\tau_{eq}$ and is independent of the gas inflow rate $\Phi$ and of the individual values of star-formation efficiency $\varepsilon$ and mass-loading factor $\lambda$. To understand the action of $\tau_{eq}$ in regulating the sSFR evolution better, in the left panel of Figure 3 we plot the sSFR histories determined from exact numeric solutions for different values of $\tau_{eq}$. For simplicity, we set R=0, then sSFR$_{int}$= sSFR. As discussed before, a non-zero R will increase the amplitude of the sSFR($t$) by a small factor of 1/(1-R) and maintain its logarithmic shape.

Interestingly, despite having changed the value of $\tau_{eq}$ by seven orders of magnitude, the resulting sSFR histories differ by less than a factor of four at any epoch. Since the sSFR history in the gas regulator model generally follows the specific accretion history of the DM halo, i.e. sMIR$_{DM}$, when $\tau_{eq}$ is very small (e.g. $\tau_{eq}$ < 0.1 Gyr), the sSFR becomes almost the same as the sMIR$_{DM}$. This is the lower limit of the sSFR history (with R=0), which is clearly shown in the left panel of Figure 3 that the red curve ($\tau_{eq}$ = 0.1 Gyr) almost completely overlaps with the green curve which represents the sMIR$_{DM}$. When $\tau_{eq}$ becomes very large, the sSFR evidently saturates at some upper limit of the sSFR history. This is seen in the left panel of Figure 3 that the purple curve ($\tau_{eq}$ = 10$^3$ Gyr) is overlapped with the black curve ($\tau_{eq}$ = 10$^6$ Gyr). The allowed range of the sSFR history (with R=0) for any values of $\tau_{eq}$ is shown as the grey shaded area in the right panel of Figure 3.

The existence of the lower limit and upper limit of the sSFR history can also be derived analytically as follows. From Equation (23), at a given epoch the value of the sSFR is proportional to $\tau_{eq}$. When $\tau_{eq}$ becomes very small, i.e. $t \gg \tau_{eq}$, the (minimum) lower limit of the sSFR history is given by

$$sSFR_{min}(t) \sim \frac{1}{1-R}\frac{1}{t} \qquad (50)$$

At a given epoch, the value of sSFR is at its maximum when $\tau_{eq}$ is very large. When $t \ll \tau_{eq}$, we can first do the Taylor expansion of 1- $e^{-t/\tau_{eq}}$ to the second order as $1-e^{-t/\tau_{eq}} \sim t/\tau_{eq} - 0.5\ast (t/\tau_{eq})^2$ and insert it to Equation (23). The upper limit of the sSFR history is given by

$$sSFR_{max}(t) \sim \frac{1}{1-R}\frac{2}{t} \qquad (51)$$

Therefore, for any given values of $\tau_{eq}$, the predicted sSFR histories from the analytic semi-exact solutions can only differ by a maximum of factor two at any epoch. The predicted maximum sSFR from Equation (51) is plotted as the brown dashed line in the right panel of Figure 3 and the predicted minimum sSFR from Equation (50) is plotted as the brown solid line in the same panel. It should be noted that the maximum and minimum sSFR derived from the analytic semi-exact solutions (i.e. from Equation (50) and (51)) are slightly different from the exact numeric solutions which are the upper and lower boundaries of the grey shaded area in the same plot. This is because, as discussed before, the analytical semi-exact solutions are derived under the assumption that the gas inflow rate is constant or only changes slowly with time which is not true in reality, especially at z > ~3. However, as we have stressed before, the analytical semi-exact solutions are very good approximations to the exact numeric solutions and more importantly, they have clearly outlined the dynamical behaviors of the exact numeric solutions, such as the reason why there should exist a lower limit and an upper limit of the sSFR histories.

Since the values of $\varepsilon$ and $\lambda$ are expected to depend on stellar mass (or halo mass), so does $\tau_{eq}$. However, as discussed above, the sSFR history in the gas regulator model is rather insensitive to the value of $\tau_{eq}$. The maximum and minimum sSFR only differ by less than a factor of four (determined from exact numeric solutions) at any epoch. This in fact naturally explains that even $\tau_{eq}$ itself may strongly depend on stellar mass, there is only a weak dependence of the sSFR on stellar mass, as observed (e.g. Elbaz et al. 2007; Daddi et al. 2007; Dunne et al. 2009; Peng et al. 2010; Rodighiero et al 2011). We will show in Section 5 that since $\tau_{eq}$ is estimated to be shorter for more massive galaxies, at a given epoch the sSFR of massive galaxies will be lower than that of the low mass galaxies. In other words, the logarithmic slope of the sSFR-$M_{star}$ relation should be negative, as observed.

It should be noted that, although we find in the gas regulator model that the difference between the minimum and maximum sSFR is less than a factor of four at any epoch for any values of $\tau_{eq}$, it does not mean the scatter of the observed sSFR should also be smaller than a factor of four. This is because we deal with an ideal situation here by assuming the gas inflow to be smooth with time. In reality, the gas inflow is likely to be clumpy and if the galaxy accretes a large amount of gas in a very short timescale, the gas mass of the galaxy increases instantaneously which will then trigger starburst and boost the SFR. The stellar mass will respond to the gas change (and SFR change) at a delayed pace, as the stellar mass is the integral of the SFR history. Therefore in this case the sSFR can change rapidly in a short timescale and the scatter of the sSFR, i.e. the deviation of the sSFR of the galaxy from the average sSFR of the galaxy population, can be larger than a factor of four. Alternatively, a sudden change of the star formation efficiency (for whatever reasons, e.g. feedback) can also result in a rapid change of the SFR and thus of the sSFR. However, since the gas inflow rate and star formation efficiency averaged over the galaxy population are expected to vary smoothly with time, the predicted sSFR evolution from the gas regulator model as shown here can be safely applied to interpret the average sSFR history of the galaxy population. For the same reason, when applying the gas regulator model to study the evolution of individual galaxies over a short timescale, or galaxies in some transient phase such as the starburst galaxy or galaxy in the process of being quenched, one should not generally assume the gas inflow, the star-formation efficiency or/and the mass-loading factor to be constant or only change smoothly with time.

**4.2 The difficulty to reproduce the observed sSFR**

In the right panel of Figure 3, we plot the sMIR$_{DM}$ multiplied by a factor of 2 (green dashed line), which is a good representation of the predicted sSFR histories from typical SAMs (e.g. Weinmann et al. 2011, Davé et al. 2011a). Interestingly, this is nicely consistent with the typical sSFR histories predicted from the gas regulator model, as the green dashed line lies roughly in the middle of the (narrow) grey shaded area. This suggests that, despite of its simple formulation, the gas regulator model has indeed captured the main mechanisms that control the evolution of the sSFR as implemented in a more complicated and realistic way in the typical SAMs.

In Figure 4 we show the observed sSFR summarized from various works in the literature. The high redshift measurements are nebular emission line corrected values. The black line and the grey line are predictions from the improved gas regulator model that will be presented in the next paper, plotted here as good representations to the observed sSFR history. Since in some of the literature it is not clear how the stellar masses are defined, we show the predicted sSFR histories for both stellar mass definitions, i.e. both as the actual stellar mass in the galaxy (i.e. $M_{star}$, in grey line) and as the integration of the SFR-history (i.e. $M_{star,int}$, in black line). As discussed above, the sSFR history following the two definitions of stellar mass differs by a factor of (1-R).

Since the sSFR histories shown in Figure 3 are computed by assuming R=0 in the gas regulator model, we take the black line in Figure 4 as a good presentation to the observed sSFR history and plotted as the black solid line in the right panel of Figure 3. Again, a non-zero R will only increase the amplitude of the sSFR($t$) by a factor of 1/(1-R) and maintain its logarithmic shape. Clearly, the observed sSFR history is fundamentally different from the predicted sSFR from both typical SAMs and gas-regulator model. The predicted sSFR histories from typical SAMs and gas regulator model have a very smooth shape that mimics the shape of the $sMIR_{DM}$. In contrast, the observed sSFR has a clear turnover around z~2, i.e. around the peak of the cosmic star formation history. It appears that the observed sSFR history is governed by different physics before and after z~2. The predicted sSFR histories from typical SAMs and gas regulator model have also underestimated the sSFR at z~2 by almost an order of magnitude. The fundamental differences in the shape of the sSFR history and the amplitude of the sSFR at z~2 have hence strongly suggested that some key process(es) is(are) missing in both typical SAMs and gas regulator model.

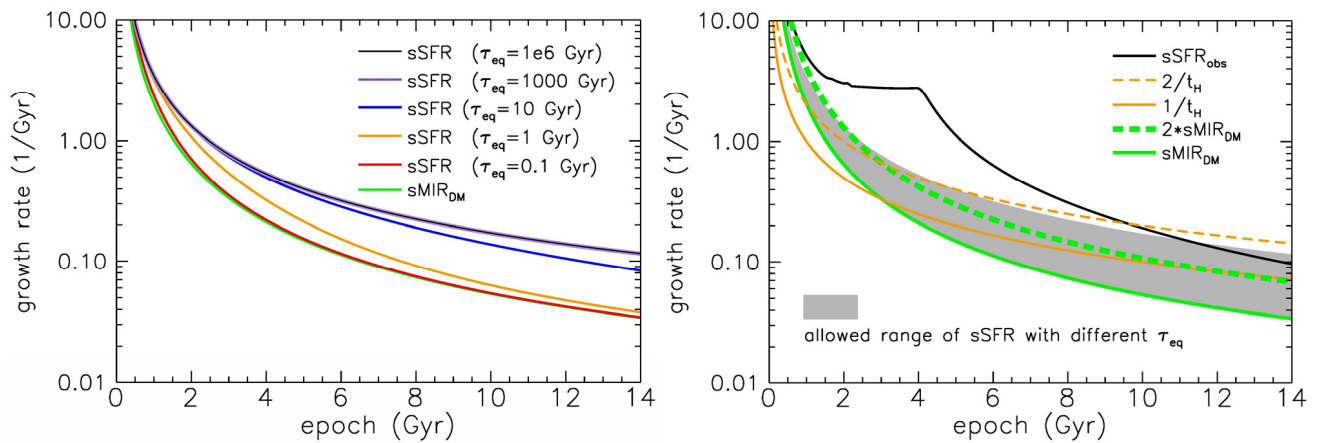

Figure 3. Left panel: The sSFR histories determined from exact numeric solutions for different values of $\tau_{eq}$. Despite the value of $\tau_{eq}$ has changed by seven orders of magnitude, the resulting sSFR histories only differ by less than a factor of four at any epoch.   Right panel: The minimum (brown solid line) and maximum (brown dashed line) sSFR history derived from the analytic semi-exact solutions as in the text. The allowed range of the sSFR history determined from exact numeric solutions (taken from left panel) is shown as the grey shaded area. The green dashed line shows $2*sMIR_{DM}$ which is a good representation of the predicted sSFR histories from typical SAMs (e.g. Weinmann et al. 2011, Davé et al. 2011a). The black line shows the representation of the observed sSFR-history as described in the text and in Figure 4. Evidently, the predicted sSFR history from the gas regulator model is in good agreement with the prediction from typical SAMs, as the green dashed line lies roughly in the middle of the grey shaded area. However, the observed sSFR history is fundamentally different from the predicted sSFR history from both typical SAMs and gas regulator model. This implies that some key process is missing in both SAMs and gas regulator model.

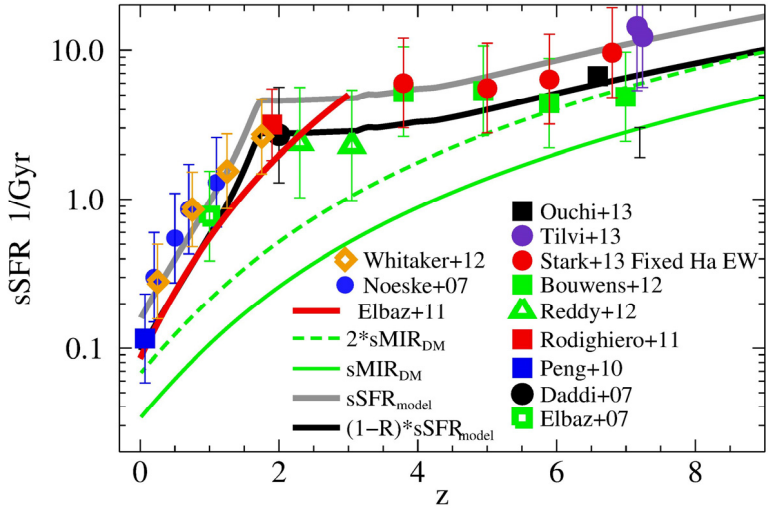

Figure 4. Evolution in the specific SFR for star forming main sequence galaxies at stellar mass $M_{star} \sim 5.0 \times 10^9 M_\odot$. Data are taken from various literatures as shown in the label. Particularly, the green solid squares show the dust corrected values from Bouwens et al. (2012). The red dots, purple dots and black square show the nebular emission line corrected values from Stark et al. (2013), Tilvi et al. (2013) and Ouchi et al. (2013), respectively. The red line shows the fitting function of sSFR = 26 $t^{-2.2}$ Gyr$^{-1}$ up to z ~ 3 from Elbaz et al. (2011). The green solid line and green dashed line show sMIR$_{DM}$ and 2*sMIR$_{DM}$ as in Figure 3 for references. The black line and the grey line are predictions from the improved gas regulator model (that will be presented in the next paper), plotted here as good representations to the observed sSFR history. Since in some of the literature it is not clear how the stellar masses are defined, we show predictions for both stellar mass definitions, i.e. both as the actual stellar mass (grey line) in the galaxy and as the integration of the SFR-history (black line). The sSFR history following the two definitions of stellar mass differs by a factor of (1-R), where R is the mass return fraction from stars.

Finally, in our analysis above, we have assumed both ε and λ, thus $\tau_{eq}$, to be constant with time. Is it possible to reproduce the *observed* sSFR history by using evolving ε and λ? If the change of ε and λ is slow, i.e. the timescale of the change is longer than $\tau_{eq}$, our conclusions above remain unchanged. In fact, even if the change of ε and λ is fast, but not in a catastrophic way (e.g. change exponentially) for an extended period of time, the range of the allowed sSFR history remains similar to the grey shaded area in the right panel of Figure 3, due to the existence of the upper limit and lower limit of the sSFR-history for any (constant) values of ε and λ.

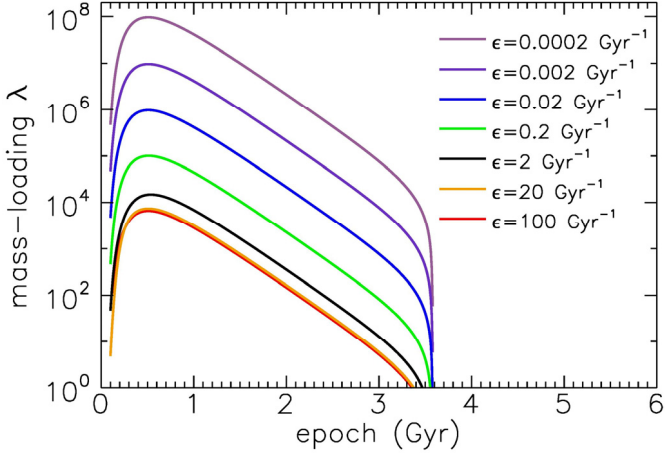

Figure 5. The required mass loading factor λ in the gas regulator model to reproduce the observed sSFR history as a function of epoch for different value of the star formation efficiency ε. A tremendous mass-loading factor, which is too large to be physically realistic, is required in the first two or three billion years to suppress the early star formation.

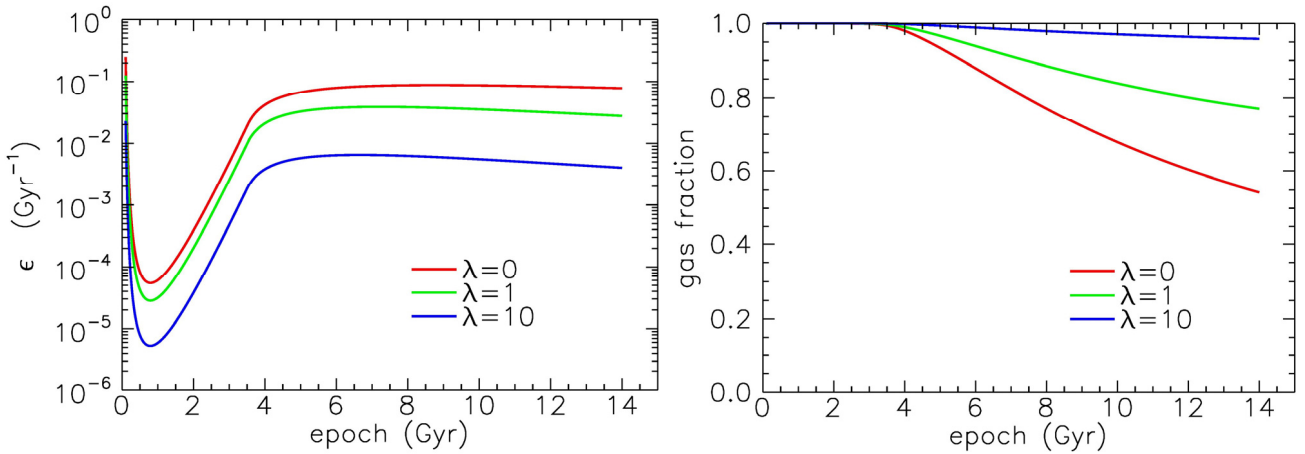

Figure 6. The required star formation efficiency ε (left panel) and the associated gas faction (right panel) in the gas regulator model to reproduce the observed sSFR history as a function of epoch for different value of the mass-loading factor λ. As the direct consequence of the small value of ε that is required to suppress the early star formation in order to reproduce the observed sSFR, the associated gas fraction is almost 100% at $z>\sim2$ ($t<\sim3.5$ Gyr) for all realistic values of λ, which clearly contradicts to the observed gas faction at similar redshifts.

Now, if we assume that ε and λ can change arbitrarily in a catastrophic way, will the gas regulator model be able to reproduce the observed sSFR history? In Figure 5, we show the required λ to reproduce the *observed* sSFR history as a function of epoch for different values of ε. Indeed, if λ can change in an exponential way, the gas regulator model is able to reproduce the *observed* sSFR history mathematically. However, the required values of λ are far too large to be physically realistic. Such a tremendous mass-loading factor is required in the first two or three billion years because the galaxy has accreted too much gas in the beginning due to the strong inflow. Without high mass loading, the galaxy has formed too many stars. This extra stellar mass formed at high redshifts will be carried on by the galaxy to later epochs and contribute to lower down the sSFR at lower redshifts, as the stellar mass is the denominator in the sSFR definition. In other words, *the main reason why the current gas regulator model has largely underestimated the sSFR at around z~2* *and thus cannot fully reproduce the observed sSFR history (see Figure 4) is not because it has underestimated the SFR at z~2, but is because it has overproduced too many stars at higher redshifts of z>2*. We suspect, that the typical SAMs have similar problems in reproducing the observed sSFR history for the same reason. Therefore the galaxy needs to find a way to get rid of this extra gas and/or prevent it from forming stars in the first two or three billion years in order to reproduce the observed sSFR history.

One way is to use an unrealistically large mass-loading factor $\lambda$, i.e. a very strong outflow as shown in Figure 5. Alternatively, the early star formation can be suppressed by tuning down the star-formation efficiency $\varepsilon$. In the left panel of Figure 6 we show the required $\varepsilon$ in the gas regulator model to reproduce the observed typical sSFR history as a function of epoch for different values of $\lambda$. The shape of the required $\varepsilon$ evolution mirrors that of the required $\lambda$ evolution as shown in Figure 5. This is because, as discussed before, $\varepsilon$ and $\lambda$ are completely degenerate in controlling the sSFR evolution and modifying $\varepsilon$ is fully equivalent to modifying $\lambda$. In the right panel of Figure 6 we show the associated gas fraction evolution for different values of $\lambda$. It is evident that the gas fraction of the galaxy is almost 100% at z>~2 (i.e. $t$<~3.5 Gyr) for realistic values of $\lambda$, which clearly contradicts to the observed gas faction at z>~2 (e.g. Tacconi et al. 2013; Troncoso et al. 2014). This is the direct consequence of the small value of $\varepsilon$ that is required to suppress the early star formation in order to reproduce the observed sSFR. Therefore, even the required values of $\varepsilon$ as shown in the left panel of Figure 6 might seem to be plausible, the associated high gas fraction (~100%) makes it unlikely to be a physically meaningful option to reproduce the observed sSFR.

Is there any other possibility in the gas regulator model to reproduce the observed sSFR-history? As discussed before in Section 4.1, under certain assumptions the typical gas regulator is incapable of reproducing the observed sSFR-history. These assumptions are that $\Phi$, R, $\lambda$ and $\varepsilon$ are constant or the timescales of the change are longer than the equilibrium timescale $\tau_{eq}$. Therefore, the key to reproduce the observed sSFR history is to violate these assumptions, by using fast evolving $\Phi$, R, $\lambda$ and/or $\varepsilon$. We have just showed above that using fast evolving $\lambda$ and $\varepsilon$ are indeed able to reproduce the observed sSFR history mathematically, but at the cost of unrealistically large mass-loading factor or high gas fraction (~100%) that contradicts to observations. Since the mass return fraction R is unlikely to evolve rapidly with epoch, the only possibility left to reproduce the observed sSFR history in the typical gas regulator mode is that the gas inflow rate of the galaxy $\Phi$ has to evolve rapidly with epoch.

According to Equation (3), $\Phi$ is given by the product of $f_{gal}$ and the gas accretion rate of the halo, i.e. $f_b*dM_{halo}/dt$, where $f_b$ = 0.155 is cosmic baryon fraction and $f_{gal}$ is the fraction of incoming baryons that flow from the surroundings into the halo and then penetrate down to enter the galaxy as baryonic gas. Although the gas accretion rate of the halo does increase exponentially in the first one billion years (top left panel in Figure 1), the shape of the resulting sSFR history, by assuming constant $f_{gal}$, simply mimics that of the specific accretion rate of the haloes and is fundamentally different from the observed sSFR history as shown in Figure 3. As discussed in Section 4.1, the normalization of the sSFR history can be easily adjusted by tuning $\tau_{eq}$, we hence stress that it is important for the model to also reproduce the characteristic shape of the observed sSFR history, especially the observed kink at the peak of the cosmic star formation history at z~2 as shown in Figure 4.

We therefore conclude that in the typical gas regulator model, the key to reproduce the observed sSFR history, especially its characteristic shape including the kink at z~2, is to find the required functional format of the $f_{gal}$. It is important to note that in this context, the meaning of the $f_{gal}$ will have to change accordingly. This means $f_{gal}$ does not only includes the effect of the penetration efficiency which has a typical constant value of order ~ 0.5 as discussed in Section 2.1, it also includes any physical mechanism that would act to influence and change the gas accretion process of the galaxy, such as feedback, gas cooling, ram-pressure stripping, a possible minimum halo mass threshold on gas accretion (Bouché et al. 2010) and etc. In fact, in terms of suppressing star formation, reducing the gas accretion of the galaxy is effectively equivalent to tuning down the star formation efficiency. However, the important difference is that reducing the gas accretion will keep the gas outside the galaxy and hence it will not contribute to increase the gas fraction of the galaxy. While keeping the gas inside the galaxy and tuning down the star formation efficiency will then cause the high gas fraction (~100%) problem as shown in Figure 6.

However, it is not trivial to identify the required functional format of $f_{gal}$, since the goal is not only to reproduce the observed sSFR history, but also to reproduce simultaneously other observed key features of the galaxy population, including distribution functions such as the stellar mass function, and scaling relations such as mass-metallicity relation, fundamental metallicity relation (FMR), gas fraction, stellar mass to halo mass ratio etc., as most of the galaxy properties are either directly controlled by or closely related to the sSFR. We will continue to explore the required functional format of the $f_{gal}$ and the implied physical mechanism in the next paper in this series.

## 5. THE CRITICAL ROLE OF $\tau_{eq}$

In many gas regulator model (or bathtub-model) papers, it is usually assumed that all galaxies live around their equilibrium state and all the analyses are based on this critical presumption. In Section 2 we have shown that the timescale for the galaxy to reach its equilibrium state is $\tau_{eq}$. $\varepsilon$ and $\lambda$ are completely degenerate in determining $\tau_{eq}$. Although the analytical form of $\varepsilon$ and $\lambda$, thus the form of $\tau_{eq}$, cannot be determined in the current paper, it is possible to estimate its orders of magnitude from observations. As Equation (7) in Lilly et al. (2013), from the definition of sSFR and Equation (4), it is easy to derive that

$$\varepsilon = sSFR \cdot \frac{M_{star}}{M_{gas}} \qquad (52)$$

Inserting Equation (52) into (12) gives

$$\tau_{eq} = \frac{1}{sSFR \cdot (1 - R + \lambda)} \frac{f_{gas}}{1 - f_{gas}} \qquad (53)$$

In principle all the quantities on the right hand side of Equation (53) can be observationally determined. The mass-loading factor $\lambda$ is typically of order unity as found in simulations (Davè et al. 2011b; Hopkins et al. 2012) and observations locally (Cicone et al. 2014) and at z~2 (Newman et al. 2012). $\lambda$ is also predicted to be dependent on stellar mass (or halo mass) that $\lambda \propto M_{star}^{-1/3}$ for a momentum-driven wind model (Murray et al. (2005) and $\lambda \propto M_{star}^{-2/3}$ for an energy-driven wind model (e.g., Dekel & Silk 1986). Therefore the typical $\lambda$ is expected to be smaller than unity for massive galaxy and of a few for low mass galaxies.

At z~3, massive galaxies ($M_{star} > $ ~$10^{11} M_{\odot}$) are observed to have a low gas fraction of $f_{gas}$~10% (e.g. Troncoso et al. 2014) and the sSFR is usually measured to be ~ 3 Gyr$^{-1}$ (see Figure 4). $\lambda$ is typically of order unity and we simply assume $\lambda$~1 and R~0.4. Then from Equation (53), for massive galaxies at z~3, $\tau_{eq}$~ 0.02 Gyr, which is much shorter than the Hubble timescale at z~3. In other works the gas fraction has been measured to have a higher value of ~40% for massive galaxies at z~2 (e.g. Tacconi et al. 2013). This implies a longer equilibrium timescale of $\tau_{eq}$~ 0.14 Gyr, which is still much shorter than the Hubble timescale at z~2.

In the local universe, the gas fraction of massive galaxies ($M_{star} > \sim 10^{11} M_\odot$) is measured to be $\sim 5\%$ or less (e.g. compilation of observations in Baldry et al. 2008 and Peeples & Shankar 2011; Santini et al. 2014; Boselli et al. 2014) and the sSFR drops below 0.1 Gyr$^{-1}$ (see Figure 4). If we take $f_{gas} \sim 5\%$ and sSFR$\sim$0.1 Gyr$^{-1}$, it gives $\tau_{eq} \sim$ 0.33 Gyr, which again is much shorter than the Hubble timescale at z$\sim$0. Although the equilibrium values of the galaxy properties listed in Table 1 are all evolving with time, $\tau_{eq}$ is much shorter than the Hubble timescale and hence the actual values of these galaxy properties actually can change fast enough to catch up with the (evolving) equilibrium values. Therefore, massive galaxies are likely to live around the equilibrium state over most of the cosmic time. We can simply use the equilibrium values in Table 1 (i.e. the equilibrium solution) to describe the evolution of the galaxy properties, without the need of the time decay term of $e^{\wedge}(-t/\tau_{eq})$.

Turning to the low mass galaxies ($M_{star} \sim 10^9 M_\odot$) at z$\sim$3, their gas fraction are measured to be $\sim$90% (Troncoso et al. 2014) and the sSFR is usually estimated to be $\sim$ 3 Gyr$^{-1}$ (see Figure 4). This gives $\tau_{eq} \sim$ 1.88 Gyr, which is comparable to the Hubble timescale at z$\sim$3. For dwarf galaxies ($M_{star} < \sim 10^8 M_\odot$) that are expected to be dominant by gas and have a gas fraction even higher than 90%, $\tau_{eq}$ will be longer or much longer than the Hubble timescale. In the local universe, the sSFR for low mass galaxies ($M_{star} \sim 10^9 M_\odot$) drops to $\sim$ 0.1 Gyr$^{-1}$ (see Figure 4) and the gas fraction is measured to be $\sim$60% (e.g. compilation of observations in Baldry et al. 2008 and Peeples & Shankar 2011; Boselli et al. 2014). This gives $\tau_{eq} \sim$ 9.4 Gyr, which is comparable to the Hubble timescale at z$\sim$0. For gas dominated dwarf galaxies with a higher gas fraction, $\tau_{eq}$ is even higher; for instance $f_{gas} \sim$ 80% gives $\tau_{eq} \sim$ 25 Gyr. Again, the equilibrium values of the galaxy properties listed in Table 1 are all evolving with time. For low mass galaxies and dwarf galaxies, their $\tau_{eq}$ is comparable to or longer than the Hubble timescale and this hence suggests that the actual values of the properties for these low mass galaxies will not be able to change fast enough to catch up the evolving equilibrium values. Therefore, low mass galaxies and dwarf galaxies are very unlikely to live around the equilibrium state at any epoch. We should use the full analytic solutions with the time decay term $e^{\wedge}(-t/\tau_{eq})$ in Table 1 (i.e. the semi-exact solution) to describe the evolution of these galaxy properties.

We take the plot in the bottom left panel of Figure 1 as a simple example. In that plot we show the gas metallicity evolution determined from the exact numeric solution in black solid curve and from the analytic semi-exact solution given by Equation (35) in black dashed curve. The equilibrium metallicity given by Equation (36) is plotted as the horizontal red solid line. In this case $\tau_{eq}$=10 Gyr and is thus longer than the Hubble timescale at early epochs and comparable to the Hubble timescale at late epochs. This plot clearly shows that the actual gas metallicity of the galaxy is evolving towards the equilibrium value, but never reaches it. Therefore, for a galaxy with a long $\tau_{eq}$, the actual gas metallicity is not only controlled by the value of the equilibrium metallicity but also critically by the equilibrium timescale $\tau_{eq}$.

One might argue that the low mass galaxies and dwarf galaxies contribute only a small fraction to the total stellar mass/light in the local universe and thus are not of great importance. However, one should also note that the star-forming galaxies on the star-forming main sequence can increase their stellar mass, if not been quenched earlier, by a factor of 30 from z$\sim$2 to z$\sim$0 and by a factor of more than 200 from z$\sim$3 to z$\sim$0 (Renzini et al. 2009; Peng et al. 2010). Low mass galaxies ($M_{star} \sim 10^9 M_\odot$) at z$\sim$3 are expected to be the progenitors of today's Milky Way-like galaxies (i.e. L* galaxies). Therefore, today's typical L* galaxies may live around the equilibrium state locally, but when we trace them back to earlier epochs, they are likely to be completely out of the equilibrium. This therefore emphasizes the important role of $\tau_{eq}$ in controlling the evolution of galaxy, and that one should not simply assume all galaxies to live around the equilibrium state at all epochs and then apply the simple equilibrium solutions (without the term of $e^{\wedge}(-t/\tau_{eq})$) to study the evolution of the properties of the galaxy population such as SFR, $M_{star}$, $M_{gas}$, $Z_{gas}$, as well as other related observational phenomenon such as the intrinsic scatters of the SFR-$M_{star}$ relation and the fundamental metallicity relation (FMR, Mannucci et al. 2010).

# 6. SUMMARY

In this paper, we extend the studies of the typical gas regulator model (or bathtub-model) as in Lilly et al. (2013) in a more thorough and self-consistent way. Our main goal has been to explore the dynamics of the gas regulator model and to study the evolution of the key properties of the galaxy population in the model, such as SFR, specific SFR, stellar mass, gas mass, metallicity, etc. We emphasize the critical role of the equilibrium timescale $\tau_{eq}$ in the gas regulator model in controlling the evolution of the galaxy population, which has been largely ignored in many similar studies. In particular, we find with the current implementation of the gas regulator model, it is very difficult to reproduce the observed sSFR-history and additional physical process is required. The main results of the paper may be summarized as follows.

1. The typical gas regulator model is based on five basic equations, i.e. Equation (4)(5)(7)(8) and (9). Equation (4) is the star formation law and Equation (5) links the wind outflow to the star formation rate. Equation (4) and (5) can also be regarded as the definition of the star formation efficiency and mass-loading factor. Equation (7)(8)(9) describes the change in stellar mass, gas mass and metal mass respectively. From these five simple equations, we derive the general analytic form of the key galaxy properties, i.e. $M_{gas}(t)$, SFR($t$), $M_{star}(t)$, sSFR($t$), $f_{gas}(t)$, $Z_{gas}(t)$, $Z_{star}(t)$, $f_{star}(t)$, $f_{out}(t)$ and $f_{res}(t)$, as summarized in Table 1. Different from other similar presentations, which investigate only the equilibrium solutions of these quantities, the ones presented in this work are the general analytic solutions, without the assumption that galaxies live in the equilibrium state.

2. The equilibrium timescale $\tau_{eq}$ is the central parameter in the gas regulator model that is in control of the evolution of all the key galaxy properties as listed in Table 1. The star-formation efficiency $\varepsilon$ and mass-loading factor $\lambda$ are fully equivalent and degenerate in determining $\tau_{eq}$. In terms of $\tau_{eq}$, we are able to write the analytic solution of the key galaxy properties in very simple and symmetrical forms as summarized in Table 1. This then enables us to easily interpret the dynamical behaviors of the evolution of these galaxy properties (as summarized in Section 2.3).

3. Since $\tau_{eq}$ is the equilibrium timescale, i.e. the timescale for a galaxy to return to its equilibrium state from a perturbation or from an (arbitrary) initial condition, for all the key galaxy properties listed in Table 1, the scatters in most of the key scaling relations, such as the $M_{star}$-SFR relation, $M_{star}$-sSFR relation, $M_{star}$-$Z_{gas}$ relation, $M_{star}$-$f_{gas}$ relation, are all primarily governed by $\tau_{eq}$.

4. Most strikingly, apart from R (which is the mass return fraction from stars) and cosmic time $t$, the sSFR evolution is solely controlled by $\tau_{eq}$, independent of the inflow rate $\Phi$ and the individual values of the star-formation efficiency $\varepsilon$ and mass-loading factor $\lambda$. $\varepsilon$ and $\lambda$ are completely equivalent in controlling the sSFR evolution. In first approximation, by assuming the equilibrium condition, sSFR is simply the reverse of the Hubble timescale (for R=0). A non-zero R will increase the amplitude of the sSFR($t$) by a small factor of 1/(1-R) and maintain its logarithmic shape.

5. Although the precise evolution of sSFR depends on $\tau_{eq}$, in fact the sSFR history is broadly insensitive to the values of $\tau_{eq}$. The shape of the sSFR history simply mimics that of the specific accretion rate of the haloes, i.e. sMIR$_{DM}$, with the typical value of the sSFR around 2*sMIR$_{DM}$. The difference between the minimum and maximum sSFR at any epoch is less than a factor of four for any values of $\tau_{eq}$, even with extreme assumptions. In other words, changing the star formation and outflow processes, and/or the feedback process (which acts equivalently to modifying the star formation efficiency and mass-loading) will not change much of the sSFR history. This distinct feature of the sSFR evolution explains why in simulations it is very difficult to reproduce the observed sSFR evolution by only tuning feedback, star formation and wind processes (e.g. Weinmann et al. 2011). The insensitivity of the sSFR-history to $\tau_{eq}$ also naturally explains that even $\tau_{eq}$ itself may strongly depend on stellar mass, but there is only a weak dependence of the sSFR on stellar mass as observed. Since $\tau_{eq}$ is shorter for more massive galaxies, the sSFR of massive galaxies is lower than that of the low mass galaxies, i.e. the logarithmic slope of the sSFR-$M_{star}$ relation should be negative, as observed.

6. The predicted sSFR history from the gas regulator model is in good agreement with typical SAMs. This suggests that, despite of its simple treatment, the gas regulator model has indeed captured the main mechanisms that control the evolution of the sSFR as implemented in a more complicated and realistic manner in the SAMs. However, the observed sSFR history is fundamentally different from the predicted sSFR history from both typical SAMs and gas regulator model. This hence strongly implies that some key process(es) is(are) missing in both typical SAMs and gas regulator model.

7. The main reason why the current gas regulator model has largely underestimated the sSFR at around z~2 and thus cannot fully reproduce the observed sSFR history is not because it has underestimated the SFR at z~2, but is because it has overproduced too many stars at higher redshifts than z~2. We suspect that the typical SAMs have similar problems in reproducing the observed sSFR history for the same reason. Therefore, in order to reproduce the observed sSFR history in the model, galaxies need to find a way to suppress the star formation in the first two or three billion years. However, this cannot be done by simply tuning the mass-loading factor and/or the star-formation efficiency, as this will produce physically unrealistic results (e.g. Figure 5). Some additional mechanism will be needed.

8. The equilibrium timescale $\tau_{eq}$ can be determined observationally via Equation (53). Massive galaxies ($M_{star} > \sim 10^{11} M_\odot$ at any given epoch) with low gas fraction are estimated to have a $\tau_{eq}$ that is much shorter than the Hubble timescale both at high and low redshifts. Therefore massive galaxies are likely to live around the equilibrium state across most of the cosmic time. We can simply use the equilibrium solutions in Table 1 to describe the evolution of the galaxy population, without the need of the time decay term $e^{\wedge}(-t/\tau_{eq})$.

9. Gas rich low mass galaxies and dwarf galaxies ($M_{star} \leq \sim 10^9 M_\odot$ at any given epoch) are estimated to have a $\tau_{eq}$ that is comparable to or longer than the Hubble timescale both at high and low redshifts. Therefore, low mass galaxies and dwarf galaxies are very unlikely to live around the equilibrium state at any epoch. The full analytic semi-exact solutions with the time decay term $e^{\wedge}(-t/\tau_{eq})$ in Table 1 should be used to describe the evolution of the galaxy population.

10. Since star-forming galaxies on the star-forming main sequence can increase their stellar mass very efficiently via star formation unless quenched, low mass galaxies ($M_{star} \sim 10^9 M_\odot$) at z~3 are expected to be the progenitors of today's Milky Way-like galaxies (i.e. L* galaxies). Therefore, today's typical L* galaxies may live around the equilibrium state locally, but they are likely to be completely out of equilibrium at earlier epochs. Therefore, one should not generally assume all galaxies to live around the equilibrium at all epochs and then apply the simple equilibrium solutions to study the evolution of their properties.

The gas regulator model has provided a simple analytic framework to study the evolution of the galaxy population. It acts as the important linkage between the phenomenological models, such as the one introduced in Peng et al. (2010 & 2012), and the cosmological framework of ΛCDM paradigm. In our next paper, we will show that the observed sSFR history requires the existence of a maximum *specific* gas accretion limit, or any mechanism that works equivalently to such a specific gas accretion limit for all galaxies. In other words, the cold gas supply that the galaxy has available for star formation is not only constrained by the cooling timescale $t_{cool}$ and free-fall timescale $t_{ff}$ as in White & Frenk (1991), but is also constrained by an additional gas accretion timescale $t_{acc}$, i.e. constrained by the maximum of $t_{cool}$, $t_{ff}$ and $t_{acc}$. We will show that with this new timescale added as the only new process, the gas regulator model will be able to produce perfectly the observed sSFR history (as shown in Figure 4), largely independent of any specific implementation of the initial conditions, star formation and wind processes.

Then we will bring in other observational constraints to break the degeneracy between the star-formation efficiency and mass-loading factor and constrain their analytic forms. The main goal of this series of papers is to build a simple and self-consistent analytical framework, by implementing the P10-model into the cosmological context via the gas regulator model, to describe the formation and evolution of the galaxy population from dark matter haloes to gas content and to the stellar population.

## ACKNOWLEDGEMENTS

We thank Andrea Ferrara, Simon Lilly, Alvio Renzini, Marcella Carollo and Antonio Pipino for stimulating discussions in the development of this work.